\documentclass[onecolumn,12pt,journal,compsoc]{IEEEtran}
\IEEEoverridecommandlockouts
\usepackage{mathtools,amsmath,amssymb,amsfonts}
\usepackage[hyphens]{url}
\usepackage[hidelinks]{hyperref}
\usepackage{algorithmic}
\usepackage{graphicx}
\usepackage[font = footnotesize]{subcaption} 
\usepackage[labelsep=period,font = footnotesize]{caption}
\usepackage{textcomp}
\usepackage{color}
\usepackage[table]{xcolor}
\usepackage{tabularx}
\usepackage{verbatim}
\usepackage{multirow}
\graphicspath{ {Figures/} } 
\usepackage{scalerel}
\usepackage{array,multirow}
\usepackage{amsthm}

\ifCLASSOPTIONcompsoc
  \usepackage[nocompress]{cite}
\else
  \usepackage{cite}
\fi

\newtheorem{theorem}{Theorem}[section]

\theoremstyle{definition}
\newtheorem{definition}{Definition}[section]
\DeclareRobustCommand{\stirling}{\genfrac\{\}{0pt}{}}

\begin{document}
	
\newcommand{\sml}[1]{\scaleto{#1\mathstrut}{5.5pt}}

\title{Technical Report \\
	Analytical Modeling and Throughput Computation of Blockchain Sharding}

\author{Pourya~Soltani~and
        Farid~Ashtiani
\thanks{The authors are with the Department of Electrical Engineering, Sharif University of Technology (SUT). Tehran 11155-4363, Iran (Email: pourya.soltani@sharif.edu, ashtianimt@sharif.edu)}
\thanks{A condensed version of this technical report has been submitted as a journal paper.}
}

\IEEEtitleabstractindextext{%
\begin{abstract}	
Sharding has shown great potential to scale out blockchains. 
It divides nodes into smaller groups which allow for partial transaction processing, relaying and storage. 
Hence, instead of running one blockchain, we will run multiple blockchains in parallel, and call each one a shard. 
Sharding can be applied to address shortcomings due to compulsory duplication of three resources in blockchains, i.e., computation, communication and storage. 
The most pressing issue in blockchains today is throughput. 
Hence, usually the main focus is to shard computation which leads to concurrent transaction processing. 
In this report, we propose new queueing-theoretic models to derive the maximum throughput of sharded blockchains. 
We consider two cases, a fully sharded blockchain and a computation sharding. 
In the former nodes are exclusive to each shard in terms of their responsibilities, i.e., block production, relaying and storage. 
In the latter though, only block production is exclusive and nodes relay and store every piece of information.
We model each with a queueing network that exploits signals to account for block production as well as multi-destination cross-shard transactions. 
We make sure quasi-reversibility for every queue in our models is satisfied so that they fall into the category of product-form queueing networks. 
We then obtain a closed-form solution for the maximum stable throughput of these systems with respect to block size, block rate,  number of destinations in transactions and the number of shards. 
Comparing the results obtained from the two introduced sharding systems, we conclude that the extent of sharding in different domains plays a significant role in scalability.
\end{abstract}

}

\maketitle

\IEEEdisplaynontitleabstractindextext
\IEEEpeerreviewmaketitle

\IEEEraisesectionheading{\section{Introduction}\label{sec:introduction}}

\IEEEPARstart{T}{hroughput} in Bitcoin and Ethereum networks are way below the satisfactory levels.
Although most of the participating nodes in mentioned blockchains have evolved through time, it has not led to much improvement in scalability.
It so happens that blockchains do not scale very easily.
This stems from the well-known scalability trilemma in blockchains \cite{Ethereum2.0} which states that only two properties among decentralization, security and scalability can fully be satisfied in a system.
In blockchains today, scalability is sacrificed for the sake of the other two. 
Different solutions have been proposed to address the blockchain scalability problem \cite{ScalabilitySurvey1,ScalabilitySurvey2}.
In this report, we focus on one of the most promising solutions, i.e., sharding \cite{ShardingSurvey, SoK}.


Sharding partitions the network into small, manageable groups, called shards, that run parallel to one another.
\textcolor{black}{The compulsory duplication of three resources (i.e., communication, data storage, and computation) can now be avoided for each participating node, while these overheads must be incurred by all full nodes in traditional non-sharded blockchains.}
Consequently, grouping (sharding) can be performed in three domains, i.e., computation (block production), network (communication) and storage.
The main focus usually is to shard the computation, however, other types can also be simultaneously achieved alongside it (e.g., RapidChain \cite{RapidChain}).
In particular, computation sharding allows partial transaction processing on a single node, since now each shard is only responsible for processing the jobs within the group.

Despite the simplicity of this idea, many new problems will arise in sharding since it is being used in a decentralized system.
Main challenges usually are with respect to intra-shard consensus safety and cross-shard atomicity\footnote{In order to guarantee consistency in the whole system, either all operations in a transaction must complete or none of them.} \cite{ShardingSurvey, Nightshade}. 
Intra-shard consensus safety stems from the fact that in sharded networks, attackers can dominate a single shard more easily than dominating the whole network. Shard takeover, also called $1\%$ attack, is analogous to $51\%$ attack in non-sharded blockchains where adversary has enough resources to change the state of the system.
The other issue is related to the transactions (TXs) that target multiple shards, leading to cross-shard transactions. 
There must be a shard interoperability mechanism in order to communicate and verify the transactions that are cross-shard.
Even so, guaranteeing the system consistency is a challenge.
The occurrence of orphaned blocks as the consequence of fork resolution, can compromise the validity of system state.

Nevertheless, none of the above challenges concern us here.
In this report, we model a pre-configured sharding scheme using queueing networks (QNs) to derive its maximum stable throughput. 
Even so, there can still be many configurations for our sharding scenario. 
In particular, sharding domains play a significant role in deriving a proper model.
They define participants responsibilities and thus the measure of their engagement in any of the main tasks, i.e., block production, relaying and storage. 
These responsibilities (especially those overlapped among shards' participants) will then be used to define system characteristics of our interest.
Thus, it is imperative to know in how many domains and to what extent sharding is applied.

In this report, our main focus will be on a fully sharded scenario. 
Each node then processes, relays and stores only the information that are assigned to its exclusive shard.
We first introduce the preliminary characteristics of our sharded blockchain, upon which we present an analytical model based on a QN to best fit the description. 
We then obtain a closed-from solution for the maximum stable throughput of this system, which is the limit before system overloads and delay goes to infinity.
We also briefly introduce and examine a computation sharding scenario without any network or storage sharding. 
In this setup, all the information is broadcast to, validated and stored by every node in the system. 
Nonetheless, block production is restricted only to the information related to the corresponding shard.
Our contributions are as follows:
\begin{itemize}
	
	\item We propose a QN model to best fit the characteristics of fully sharded blockchains where each shard has its own set of distinct miners. 
	In the proposed model, we exploit multi-class negative and positive signals \cite{Miyazawa} in order to model the  block production as well as multi-destination cross-shard TXs in the blockchain.
	\item Using the proposed model, we derive the maximum stable throughput of the system with respect to block size, block rate, number of shards and the number of destinations in TXs.
	We show that though the probability of TXs to become cross-shard approaches one as the number of shards in the system increases \cite{Monoxide}, in a fully sharded blockchain, throughput growth with respect to the number of shards still converges to be linear.
	\item To further illuminate the effect of sharding domains in the modeling, we modify our proposed QN model to examine a computation sharding scenario without any network or storage sharding.
	\item Since the shared network in the computation sharding scenario can ultimately become the bottleneck, we consider a parameter limiting the load on the shared network in order to limit the fork rate, then we derive the maximum throughput satisfying the constraints.   
	In this case, the system throughput cannot grow as freely as the fully sharded blockchain scenario.
	Comparing the results obtained from the two introduced sharding scenarios, we conclude that the extent of sharding in different domains plays a significant role in scalability.
\end{itemize}

It is worth noting that there are different perspectives towards the scalability performance metrics.
Transaction throughput and transaction confirmation latency are the two most talked-about \cite{ScalabilitySurvey1}.
Nonetheless, in this report, we consider throughput as the performance metric for blockchain scalability.

The rest of this report is organized as follows: 
In Section \ref{RelWorks} we briefly review important features of a well-known asynchronous sharding scheme, Monoxide \cite{Monoxide}, as well as some of the state of the art papers which benefited from queueing concepts in their blockchain modeling and analysis.
We describe our system model in Section \ref{sysModel} which completely concentrates on a fully sharded blockchain.
We then propose an analytical model based on a queueing network in Section \ref{analyticModel} that is able to capture the behavior and characteristics of our system.
This model is restricted to the case with single-destination TXs.
The extension to the case with multi-destination TXs is presented in Section \ref{extension}.
For both cases in Sections \ref{analyticModel} and \ref{extension} we derive the maximum stable throughput.
We then introduce the computation sharding scenario and the required changes to be made to our original fully sharded model to derive its throughput in Section \ref{Mere}.
Some numerical evaluations are given in Section \ref{NumResults} and finally, this report is concluded in Section \ref{Conclusion}.

\section{Literature Review} \label{RelWorks}

There are many rich proposals in the field of blockchain sharding \cite{Ethereum2.0, Elastico, RapidChain, Monoxide, Nightshade}.
Monoxide \cite{Monoxide} is the first asynchronous sharding scheme proposed and it is quite well known.
Hence, we confine ourselves to review the key concepts of this proposal here, since it is the most similar to our line of work.
Interested readers can refer to \cite{ShardingSurvey, SoK} for more in-depth information about sharding schemes.

Monoxide partitions demands based on their issuer’s account address.
Hence, each shard is responsible for providing service for a specific set of addresses (accounts) assigned to it.  
Shards then make use of Chu-ko-nu mining, a proof of work (PoW) variant, which further helps the system to handle the shard takeover problem \cite{Monoxide}.
Chu-ko-nu mining allows miners to use a single PoW solution to create multiple blocks at different shards simultaneously (a.k.a. block batching).
Consequently, miner's mining power is amplified (multiplied) by the number of shards a miner participates in.
Fortunately, this rule doesn't apply to attackers targeting a single specific shard, since Chu-ko-nu does not allow more than one block per-shard with each PoW solution. 
Hence, even though total hash power is divided by the number of shards, taking over a shard could become as hard as its non-sharded counterpart if the average number of shards miners participate in, approaches the total number of shards. 

Monoxide follows a lock-free scheme for handling cross-shard TXs which relies on receipts (RXs).
Monoxide proposes eventual atomicity where a single cross-shard TX is decoupled into an originated TX in the local shard, and a relay TX (a.k.a. receipt) being put into the outbound transaction set.
Though, this scheme leads to higher utilization and throughput, with eventual atomicity, the consequence of fork resolution in one shard may affect the validity of relay TXs forwarded and confirmed in another shard.
Therefore, the relay TX cannot be committed in destination shard until its initiative TX is placed in enough depth of the originating shard chain, incurring additional delay for cross-shard transactions.

Unlike sharding, literature on using queueing theory in analyzing blockchain characteristics is not as plentiful as it could be.
Still, some inspiring line of research can be found in the literature.
In \cite{Kawase_2017}, the authors modeled the Bitcoin blockchain via a single server queue with batch service departure. 
Their ultimate goal was to derive the transaction confirmation time which is the time since the TX is issued and the time it has become part of a block. 
TXs arrive according to a Poisson process and service time interval in their model is general.
Nevertheless, there were some deviations between their obtained results and reports from Bitcoin performance.
They later addressed this issue in \cite{Kawase_2018} by considering exponential-type distributions for service time intervals.
Consequently, they were able to estimate the mean transaction-confirmation time more accurately.
To capture the effect of Bitcoin fees on TX confirmation times, the authors in \cite{Kasahara_2019} proposed a priority queueing model with batch departures for the block production process.
As the highlight of their work, they then managed to show that increasing the block size is not a fundamental solution for the scalability problem.

To better fit the real world scenario, the authors in \cite{BlockchainQueue} proposed a model for mining process in Bitcoin using two queues, i.e., block-generation and blockchain-building queues.
Specifically, they decomposed service time into two different exponential service stages of mining process and network latency.
This model further simplified the computations.
Using this method, they derived the average number of TXs in each queue, the average number of TXs in a block, and the average TX confirmation time under system stable condition.
The authors in \cite{BitNetModel} modeled each node in the network as a queue to capture the impact of many criteria such as node connectivity and block size on the data delivery protocols used in blockchains.
They further investigated many aspects of forks in blockchains like forking probability and the duration of the ledger inconsistency period.

\section{System Model} \label{sysModel}

We consider a generic pre-configured sharding mechanism to analyze its characteristics and behavior. 
The details about how shards are formed or how nodes are moved between them through time is out of the scope of this report.
We just need them to be working continuously and securely.
In our model, Nakamoto consensus family is used for intra-shard consensus, and similar to many works (e.g., \cite{BitNet, Kawase_2018, BlockchainQueue}), PoW block production time is considered to be an exponential distribution.
Hence, the resulting shards operate asynchronously with respect to one another. 
Nonetheless, as long as the objective is system throughput, the results also apply to the synchronous configurations as well.
In this setup, we use the words ``miners'' and ``nodes'', interchangeably, for full nodes participating in the consensus process. 
We assume all miners are honest and do not deviate from the system protocol. 

We consider an account-based sharding where jobs are assigned to a specific shard based on the address of their senders. 
In other words, each shard is responsible for giving service to a specific set of addresses assigned to it.
TXs are then distributed uniformly among shards. 
We assume that the issued TXs arrive at different shards as independent Poisson processes.
Upon arrival, they are first broadcast throughout the corresponding shard network, which are then added to the shard miners' memory pool (mempool) upon validation.
Each TX offers a fee in order to be added to a block by miners, which can further specify its quality of service (QoS) in terms of delay.
We assume fees are such that they give enough incentives to all miners present in a shard to produce blocks even in the least populated mempools.
Miners will then collect fees from TXs forming the block they had just mined and inform other participants in their shard network of the new state.

We consider a fully sharded blockchain, where each shard has its own set of distinct miners.
Each node mines only in one shard where also its relaying and storage responsibilities are restricted to that one shard.
In other words, each miner mines, relays and stores only the information that are assigned to its exclusive shard.
We consider nodes with rather the same capabilities, and shards with roughly equal number of them.
This property along with honesty gives us a symmetric architecture in the long run given that miners in different shards follow similar protocols.
There might be some transient effects due to churn (addition and removal of nodes) and random movements of miners, but we are only interested in steady state of the system and ignore such events.

Now, we have multiple identical shards working in parallel.
We still need a mechanism to address cross-shard TXs, especially since the probability of a TX to be cross-shard approaches one as the total number of shards increases in an account-based sharding \cite{Monoxide}.
Each TX originates from one account, however, it may have many recipients.
Thus, if the addresses of sender and receivers reside in different shards, more than one shard are affected.
We adopt the same mechanism as \cite{Monoxide} where cross-shard TXs are handled with the aid of RXs.
When a cross-shard TX is serviced in its originating shard, the shard produces receipts as the evidence for the validity of the transaction operations.
In particular, the originating shard will first deduct from the sender's account, takes care of local (intra-shard) transfers, then handles cross-shard transfers via RXs.
The produced RXs are then routed through some gateway nodes to their destination shards just after their containing block is in a secure depth of their originating chain.
So, the interaction among shards are through RXs and it is assumed here that the shards have direct communication links with each other.
This results in a fully connected communication graph between shards (unlike \cite{RapidChain}).


There might be multiple destinations in a cross-shard TX all pointing to the same shard other than the originating shard.
In this case, we assume that the originating shard will gather them all in a single RX targeting the shard responsible for serving the corresponding destinations.
Consequently, RXs can also have multiple destination fields, but they must all belong only to one shard.
For example, consider a system with three shards $A$, $B$ and $C$. 
Three TXs are issued in shard $A$ with destination shards as `$AA$', `$BB$' and `$BC$' where shard $A$ produces none, one and two receipts for each, respectively.
We could produce a RX corresponding to each destination field in TXs, but the adopted approach gives us better utilization and higher system throughput.

In our model, jobs and the amount of service they require are defined based on their sizes.
We assume transactions and receipts (TRXs) have the same size irrespective of the number of their destination fields. 
This assumption can even be very close to reality. 
There are fair share of instances that most of the TX size is occupied by their sender's private signature, e.g., Bitcoin before BIP141\footnote{Bitcoin Improvement Proposal 141.} update \cite{segWit}. 
Consequently, it is reasonable to state that all TXs and RXs impose the same amount of work on shards.
Note that the incurred load to the system and to shards are not alike.
A multi-destination TX imposes the same load on a shard as its single-destination counterpart, but it can incur much higher load on the system since it usually produces more receipts.

The same rule also applies to blocks in terms of size and their required service.
Blocks have a maximum size and hence can accommodate a limited number of TRXs.
As the number of TRXs in a block increases, validating it takes more time, imposing more load to its host shard network.  
Though it might seem the amount of service blocks require is equal to the ratio of their sizes to TRXs', this is not quite true.
In fact, we cannot compare the load incurred by TRXs and blocks with each other due to different priorities.
In other words, miners might prefer servicing one over the other.
Taking Bitcoin as an example, though block sizes are much larger than of TXs, their propagation delay is smaller \cite{BitcoinCharts}.
In this manner, we can only compare services of the components of the same type.

In the following sections, we first present an analytical model based on a queueing network that can best fit the system characteristics just described.
In order to do so, we make use of product-form queueing networks (PFQNs) which rid us from the complication of solving a multi-dimensional Markov chain.
Using the obtained model then, we derive the maximum stable throughput of a sharded blockchain beyond which it becomes overloaded and delay in the system goes to infinity.

It is worth noting that stability is not the only important factor in computing throughput.
In fact, stability is just a necessary condition but it’s not sufficient.
Security concerns are much more important.
Consider fork rate as an example. 
Forks cause inconsistency in blockchain view and can ultimately compromise system safety \cite{BitBackbone}.
Hence, depending on system characteristics and application, fork rate may confine system throughput to limit the experienced delay in block delivery.
We include the limitation on the network load as the parameter representing forking concern whenever the delay impact is no longer negligible.
Nonetheless, we mainly focus on stability which can further give us an upper-bound on sharding throughput.

\section{Analytical Model} \label{analyticModel}

In this section, we propose a QN model to represent what we described in the previous section.
In particular, we model each shard with a couple of queues which can further interact with other shards' queues.
In this respect, TRXs play the role of customers in the queues.
As explained in the previous section our main expectation for the model is to enable us to derive sharding throughput. 
In this regard, the main challenge in our modeling approach is due to batch movements.
Owing to block production, we have batch departure from some queues which instantly lead to batch arrivals to some other queues.
Thus, we need to know the average number of customers per batch because not all the mined blocks are full size. 
In other words, it is possible to have partial batches due to not having sufficient number of TRXs at the time of block production.

In order to compute the average size of the mined blocks we need to solve a multi-dimension Markov chain with dimensions proportional to the number of shards.
Writing global balance equations (GBE) for such a chain is itself complicated, let alone solving them.
To address this challenge, we make use of nonlinear QNs \cite{Miyazawa}.
These networks are suited to model concurrent movements of customers among queues. 
In our approach, we are also able to include partial batches.
In particular, we make use of quasi-reversible (QR) queues with signals to construct a PFQN as our analytical model.
The detail of quasi-reversibility is far beyond the scope of this report, though we just explain some of its significant characteristics in the following.

Quasi-reversibility is an input-output property of a queueing system. 
It implies that when the system is in stochastic equilibrium, the future arrivals, the current state of the system, and the past departures are independent \cite{Miyazawa}.
It equips us with the relation between departure rate and arrival rate for each queue, when we need to write traffic equations of a QN.
Traffic equations describe how arrival rates to and departure rates from queues balance with each other.
In particular, in a QN, we know only the arrival rates from the outside world and routing probabilities among queues.
We do not have sufficient knowledge about the rates routed from queues to each other.
QR enables us to derive departure rates from each queue which can further be used to obtain arrival rates to queues through traffic equations.

Furthermore, QR leads to product-form queueing networks \cite{Miyazawa}.
In this class of networks, the joint distribution of all queues is the product of the marginal distributions of the individual queues, provided that the corresponding traffic equations are satisfied. 
Due to this property, the analysis of a QN reduces to the analysis of each single queue solely, as well as solving the traffic equations \cite{Miyazawa}.
In other words, we can isolate each queue from the network and examine it individually after solving the traffic equations.

In the sequel, in Section \ref{Review}, we review the fundamental concepts of nonlinear QNs.
This review is rather due to two less known entities in these networks, positive and negative signals, and how to ensure QR property in their presence. 
In Section \ref{analyticalModel}, we try to model the system to best fit the description in Section \ref{sysModel} which we then obtain its traffic equations. 
Finally, in Section \ref{shardingThroughput} we derive the sharding throughput with regard to the obtained traffic equations.

\subsection{A Brief Review on Nonlinear Queueing Networks} \label{Review}

There are three types of entities in these networks: regular customers, negative signals and positive signals.
Regular customers are well known in queueing theory.
They are the only type of entities waiting in queues to be served.
When a negative signal arrives at a queue with $n$ customers, it causes the customer in position $l$ to leave with probability $\eta(l, n)$ given $ \sum_{l=1}^{n} \eta(l, n) = 1 $, while customers in higher positions fill in the gap.
A negative signal disappears without any effect when it arrives at an empty queue, otherwise it immediately triggers another entity (usually of its kind, i.e., a negative signal) at the queue output.  
On the other hand, a positive signal similarly triggers another entity (usually of its kind, i.e., a positive signal) at the queue output, but it increases the number of customers in the queue by one \cite{Miyazawa}.
In this report, signals always trigger their own kind at the output of their visiting queue.

Customers, positive signals and negative signals are denoted by $c$, $s^+$ and $s^-$, respectively. 
Each can further have multiple classes $i$, denoted by $c_i$, $s_i^+$ and $s_i^-$ ($i = 1, 2, ...$) with arrival rates to queue $J$ as $\alpha_{Jc_i}$, $\alpha _{Js_i}^ +$ and $\alpha _{Js_i}^ -$, respectively. 
Upon departure, each entity can change its class or even its type via network routing parameters, e.g., entity $u$ departing queue $J$ can turn into an entity of type $v$ destined for queue $K$ with probability $r_{Ju,Kv}$.
The only requirement for routing probabilities is $\sum_{K} \sum_{v} r_{Ju,Kv} = 1$ for $\forall J, u$.

\begin{figure}[t!]
	\centering
	\includegraphics[width=0.85\linewidth]{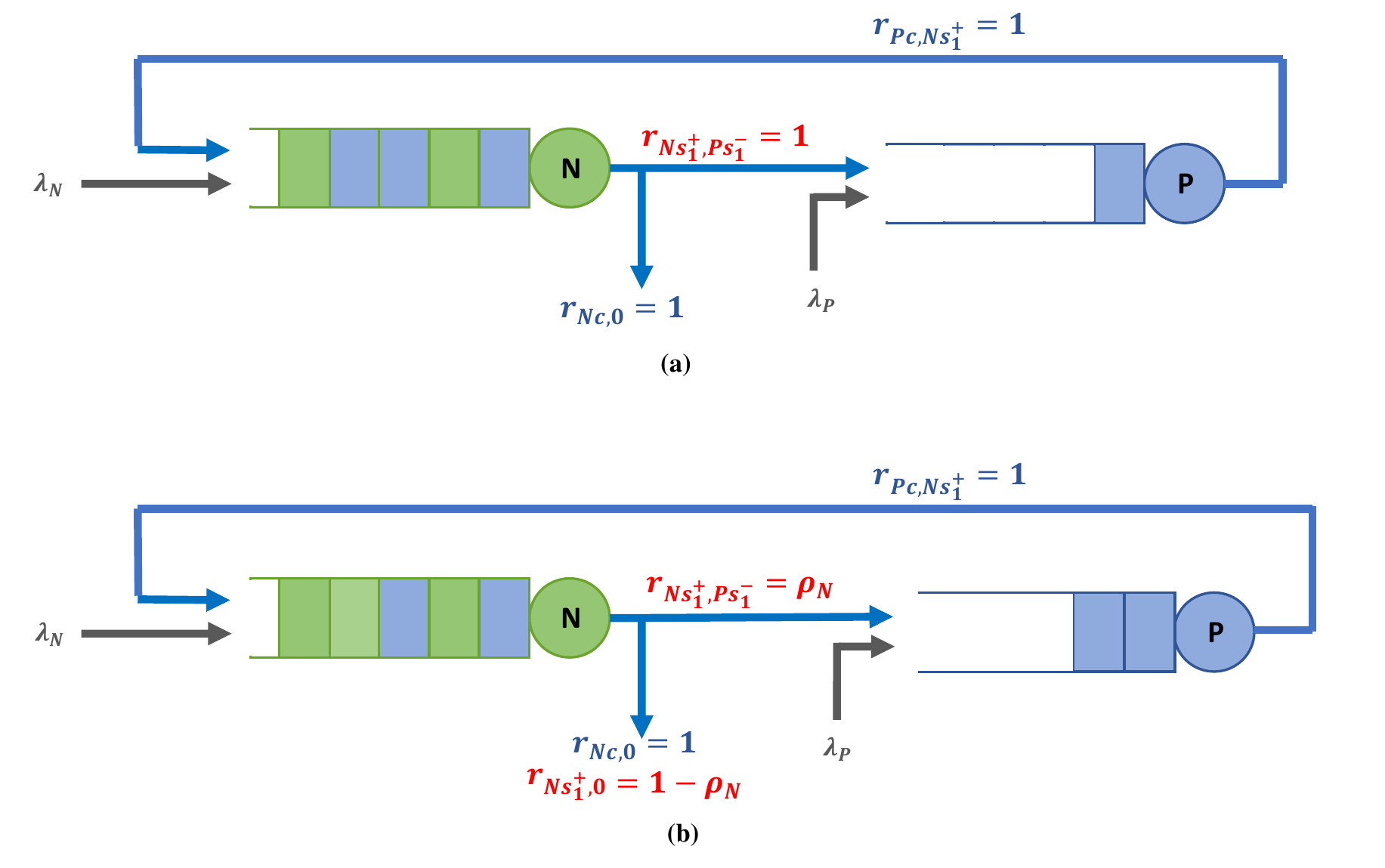}
	\caption{Using signals to model singleton and batch departures from queue $P$ entering queue $N$. (a) Without additional positive signal departure, QR property is not satisfied. (b) Including additional positive signal departure in such a way to both keep the rates unchanged and still satisfy QR property.}
	\label{fig:Example}
\end{figure}

As an example, let us consider the scenario shown in Fig. \ref{fig:Example}a where we have two $M/M/1$ queues in tandem.
Customers arrive at queue $P$ and $N$ from the outside according to Poisson processes with rates $\lambda_P$ and $\lambda_N$, respectively.
There is only one class of customer shown by $c$ in both queues. 
In queue $P$, customers are served in batches of size two whenever there are more than one customer.
Otherwise, the departure would be singleton.
The service time of both batch and singleton departures are exponentially distributed with rate $\mu_P$ and the service discipline is first-come-first-serve (FCFS). 
The departed customers (of size one or two) then join queue $N$, which also has exponentially distributed service time but with rate $\mu_N$.
In this scenario, there are times that two customers leave queue $P$ and enter queue $N$ together.
In order to model this event, we can use signals to address the batch movements.

To model batch departure from queue $P$, we could consider customers upon their service completion in this queue, and reroute them back to it as a negative signal, i.e., $r_{Pc,P{s _1^ -}} = 1$, in order to remove two customers at the same time.
However, this approach has a drawback, i.e., inability to address singleton departures due to partial batch problem.
This problem stems from the fact that negative signals when enter an empty queue, disappear without leaving any trace of its precedence. 
In our case, when there is only one customer in queue $P$, the triggered $s _1^ -$ will face an empty queue and disappear.
Thus, one customer has been serviced in queue $P$ yet we do not see any trace of it in queue $N$ due to partial batch problem.

In order to solve the problem of partial batch departures, we first route serviced customers from queue $P$ to queue $N$ as a positive signal, i.e., $r_{Pc,N{s _1^ +}} = 1$, which will then add a new customer to this queue and triggers a new $s _1^ +$ at its output.
Subsequently, we reroute the newly triggered positive signal back to queue $P$ as a negative signal, i.e., $r_{N{s _1^ +}, P{s _1^ -}} = 1$.
The empty queue $P$ and the resulting disappearance of the negative signal is analogous to singleton departures.
But, in case the negative signal does not face an empty queue $P$, it removes another customer from it, addressing the batch departure, and triggers a new $s _1^ -$ at the queue's output.
Finally, newly triggered $s _1^ -$ can enter queue $N$ as a regular customer with $r_{P{s _1^ -}, Nc} = 1$.

Through this simple example we saw how much practical signals can be, especially in modeling the concurrent movements.
Nevertheless, this example does not yet satisfy the QR property.
As mentioned, QR gives us the advantage to isolate each queue in the network and obtain their arrival and departure rates.
Hence, we cannot still write the traffic equations which we require for further performance evaluations.
In order to remedy the problem, we need to modify our QN such that whenever queue $N$ is empty, an additional Poisson departure process of positive signals is activated with rate ${\rho _{N}^ {-1}}{\alpha _{N{s _1}}^ +}$, where $\rho _{N} = (\lambda_N + \alpha _{N{s _1}}^ +)/\mu_N$ is the utilization factor (load) of queue $N$ \cite{Miyazawa}.
Then, quasi-reversibility of both queues $P$ and $N$ in our example will be satisfied.

In fact, in order to preserve the QR ‌property in a nonlinear QN with only one class of customer, those queues that accept positive signals must emit additional positive signals whenever being empty (see Appendix \ref{appendixA}).
However, there are now unwanted positive signals wondering among queues. 
Unfortunately, this can cause deviation from the original problem.
Thus, we here propose a mechanism to resolve this inconvenience. 
Since additional departures only happen when the queue is empty, we remove positive signals with the probability that the queue emanating them is empty.
In other words, we follow a probabilistic approach to decide whether keep the departed positive signals or route them to outside of the network.

To clarify, let us once again consider our example in Fig. \ref{fig:Example}b.
In this example, only queue $N$ hosts a positive signal. 
Thus, to satisfy the QR property, it must emit $s _1^ +$ with rate ${\rho _{N}^ {-1}}{\alpha _{N{s _1}}^ +}$ whenever it is empty.
Hence, some positive signals at the output of queue $N$ are due to service completion in queue $P$ and others due to additional departure rate.
We are only interested in the former, though we cannot distinguish between the two.
Nevertheless, we can readjust the rate using routing probabilities, without even disturbing the QR property.
In particular, we now set $r_{N{s _1^ +}, P{s _1^ -}} = \rho _{N}$ since $\rho _{N}$ is the probability that queue $N$ is not empty. 
Subsequently, we would have $r_{N{s _1^ +}, 0} = 1 - \rho _{N}$ where ``$0$'' represents the exogenous world.
So, we remove positive signals with the probability that the queue emanating them is empty and we keep them with the probability that the queue is busy.

This method has a downside, though.
It tampers with the batch size.
In the example above, there are times which there are more than one customer available in queue $P$, but probabilistic routing might result in a singleton departure instead of batch.
With the proposed probabilistic routing scheme there will be no guarantee on size of the departed batches.
Note that the missed departures (due to smaller batches) are compensated by the additional departures.
In fact, we might have smaller batches, but we also have additional departures which are randomly generated in the system and compensate for the missed batch segments.
Accordingly, our probabilistic routing does not interfere with our throughput computation, since rates remain untouched with respect to the real scenario.

Finally, we write the traffic equations.
In order to do so, we require departure rates from each queue, which can be obtained from the QR relation (see \eqref{QR2} in Appendix \ref{appendixA}).
In the nutshell, the departure rates for three entities $c$, $s^+$ and $s^-$ from queue $J$ can be computed as ${\rho _{Jc} \mu_{Jc}}$, ${\rho _{Ju}^ {-1}}{\alpha _{Js}^ +}$ and ${\rho _{Ju}}{\alpha _{Js}^ -}$, respectively, with $u$ denoting the class of customer which the signals add to or remove from queue $J$ \cite{Miyazawa}.
In our previous example, $u$ was equal to $c$, since there was only one class of customer in the system.
Now, using these rates along with routing probabilities, we can obtain the arrival rate of each class of entities to every queue.
For example, traffic equations for queues $P$ and $N$ in Fig. \ref{fig:Example}b would be  
\begin{subequations} \label{eqExample}
	\begin{align}
	\alpha _{\sml{Pc}} & = \lambda_P, \\
	\alpha_{\sml{Ps_{1}}}^ - & = {\rho^{\sml{-1}} _{\sml{N}}}  \alpha _{\sml{Ns_{1}}}^ +  r_{\sml{N{s _{1}^ +}},\sml{P{s _{1}^ -}}} = {\rho^{\sml{-1}} _{\sml{N}}}  \alpha _{\sml{Ns_{1}}}^ + {\rho _{\sml{N}}} = \alpha _{\sml{Ns_{1}}}^ + , \\
	\alpha _{\sml{Nc}} & = \lambda_N  + {\rho _{\sml{P}}} \alpha_{\sml{Ps_{1}}}^ - {r_{\sml{P{s_1 ^-},Nc}}} = \lambda_N  + {\rho _{\sml{P}}} \alpha_{\sml{Ps_{1}}}^ -,  \\
	\alpha _{\sml{Ns_{1}}}^ + & = \rho_{\sml{P}} {\mu _{\sml{P}}} {r_{\sml{Pc,N{s_1 ^+}}}} = \rho_{\sml{P}} {\mu _{\sml{P}}}.
	\end{align}
\end{subequations}
Also, subscript $c$ has not been shown in utilization factors due to presence of only one class of customer in the system.

Interested readers can refer to Appendix \ref{appendixA} for more information on this topic.

\begin{table}[t!]
	\centering
	\caption{List of Parameters}
	\begin{tabularx}{\linewidth}{ | l | X |}
		
		\hline
		Parameter & Description \\ \hline
		
		$M$ & Number of shards \\ \hline
		
		$\lambda$ & TX input rate to each shard \\ \hline
		
		$d$ & Number of destinations in a TX \\ \hline
		
		$D[d]$ & Probability mass function of $d$ \\ \hline
		
		$b$ & Maximum number of TRXs allowed in a block \\ \hline
		
		$\alpha _{\sml{Ju}}$ & Input rate of types $u$ customer to queue $J$ \\ \hline
		
		$\alpha _{\sml{Js_{i}}}^ -$ & Input rate of negative signal $s^- _i$ to queue $J$ \\ \hline
		
		$\alpha _{\sml{Js_{i}}}^ +$ & Input rate of positive signal $s^+ _i$ to queue $J$ \\ \hline
		
		
		$\alpha _{\sml{Jc_{i}}}^ +$ & Input rate of positive signal $c^+ _i$ to queue $J$ \\ \hline
		
		$U$ &  The largest stage possible for a signal $c^+ _i$ \\ \hline	
		
		$\stirling{n}{q}$ & Stirling number of the second kind which is the number of ways to partition a set of $n$ objects into $q$ non-empty subsets \\ \hline
		
		$R_{\sml{J'J}} ^k$ & The rate service completion of a receipt in network queue $J'$ results in a stage $k$ signal, $c_{k} ^ +$, aimed for network queue $J$ \\ \hline
		
		$\mu _{\sml{Ju}}$ & Service rate of type $u$ customer in a typical queue $J$ ($u$ is dropped if $J$ hosts only one type of customer)\\ \hline
		
		$\rho _{Ju}$ & Utilization factor or the load incurred by type $u$ customer to a typical queue $J$ \\ \hline
		
		$\rho _{J}$ & Sum of the utilization factors of all customers in $J$, i.e., $\rho _{J} = \sum_{u} \rho _{Ju}$ \\ \hline	
		
		$r _{Ju,Kv}$ & Routing probability of type $u$ entity from queue $J$ to queue $K$ as a type $v$ entity \\ \hline
		
	\end{tabularx}
	
\end{table}

\subsection{Proposed Queueing Network Model for Blockchain Sharding With Only Single-Destination Transations} \label{analyticalModel}

\begin{figure}[t!]
	\centering
	\includegraphics[width=0.90\linewidth]{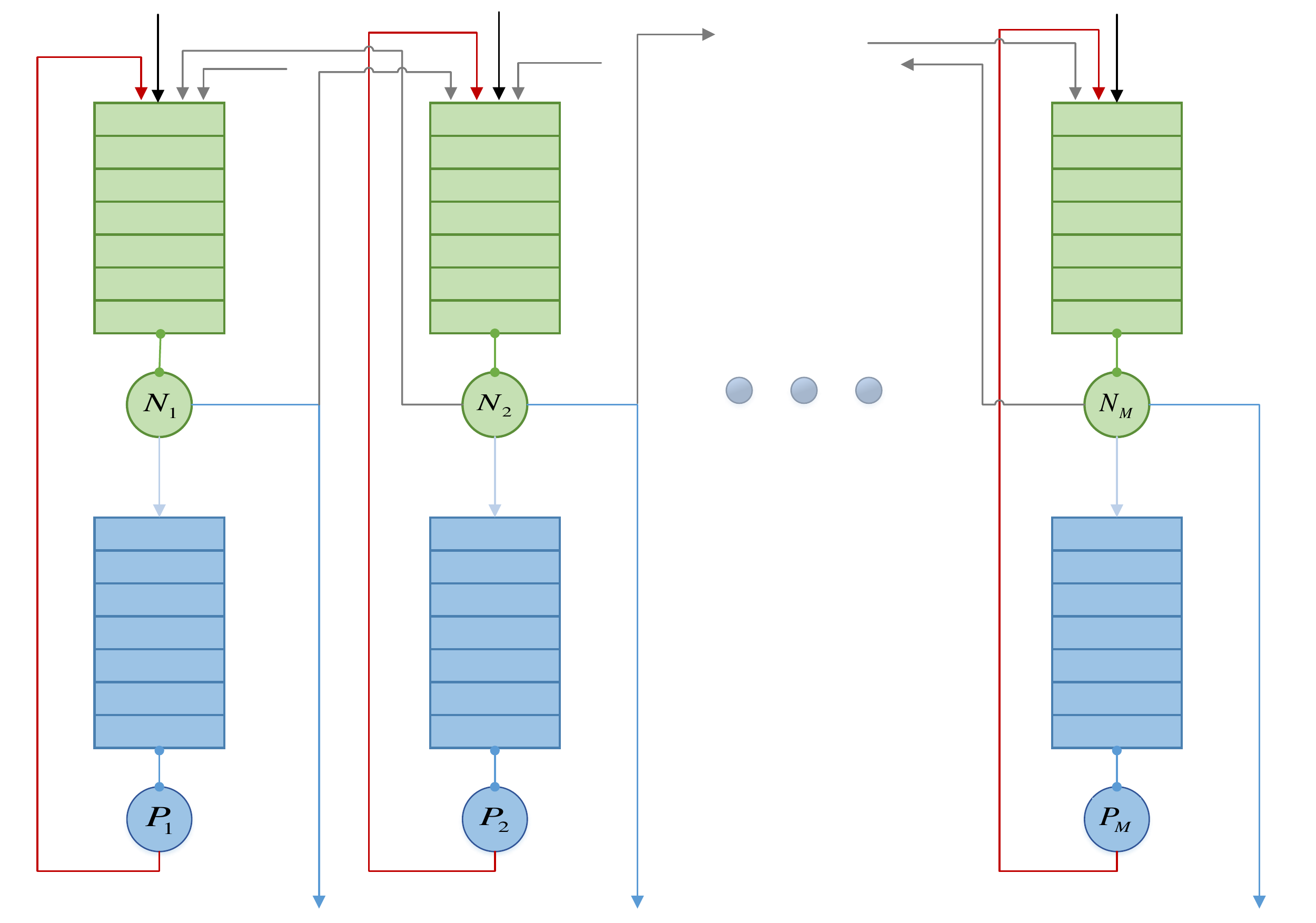}
	\caption{Sharding queueing network. transactions, receipts and blocks are illustrated with black, gray and red colors, respectively, and blue arrows show committed and finalized TRXs exiting the system.}
	\label{fig:Model1}
\end{figure}

We have a fully sharded blockchain with $M$ shards.
As a direct result of Nakamoto consensus, we model each shard with two queues, a consensus queue ``$P$'' and a network queue ``$N$'', motivated by \cite{BlockchainQueue}. 
Network queue is responsible for information dissemination in the shard and consensus queue is responsible for producing (mining) the blocks.
Fig. \ref{fig:Model1} depicts the proposed QN model which also demonstrates the interactions among shards.

In this setup, we regard $\mu_{Ju}$ as the service rate of class $u$ customer at a typical queue $J$.
$\rho_J$ is the utilization factor of queue $J$, which itself can be comprised of many classes, i.e., $\rho_J = \sum_{u}^{}\rho_{Ju}$.
Note that service rates could depend on the number of miners in each shard and accordingly on $M$.
Nonetheless, we rather ignore such dependency for the sake of simplicity and comparability.
However, one can also keep them unaffected regarding the envisioned control parameters in the protocol, e.g., mining difficulty level \cite{nakamoto2012bitcoin}.

The inputs to consensus queues are transactions and receipts.
For now, we consider them to have only one destination field.
They first need to be propagated through the shard network to be added to miners mempools.
They are then mined as part of a block and rebroadcast in the shard network to update the shard's state. 
Hence, all TRXs are first served by shard's network queue, then routed to the consensus queue to be mined.
Upon mining a block, the block is also propagated through the shard network.
In other words, the outputs of each consensus queue are directly routed to its corresponding shard's network queue.
Therefore, transactions and receipts are disseminated in the network twice, once raw (before being mined) and another time in form of a block.

Due to indifference assumption between TXs and RXs sizes (see Section \ref{sysModel}), we only consider one class of customer denoted by `$c$' arriving at consensus queues. 
In general though, TRXs might have different fees, hence they can be categorized into different classes with different priorities \cite{Kasahara_2019}. 
Nevertheless, such prioritization only matters when the objective is to obtain customers QoS, specifically TRX delays.
This is not within our objectives in this report, hence we ignore the impact of different fees.  

We consider two classes of customers arriving at network queues, `$c$' representing the TRXs just like the case in consensus queues and `$s$' representing the block components.
Each class $s$ customer might either correspond to a mined TX or RX.
Serviced RXs as part of a block in shard's network queue leave the system.
However, in case of TXs, they might turn into receipts aiming other shards.
Note that we could also add an intermediary queue to each shard right after network queues to keep RXs waiting till their source TXs are in enough depth of their shards' chain.
Nevertheless, this only incurs more delay and has no effect on throughput, hence, we disregard their existence.

The reason we are considering block components instead of block itself is mainly due to the complexity of decomposing blocks in traffic equations in order to extract RXs.
Moreover, while we cannot differ between the required service for blocks with different sizes, block components can capture the effect of block size on the required service.
Also, one might ponder the necessity to introduce a new class of customer, $s$, in the network queue while they are basically TRXs just mined.
This stems from the fact that due to priority, blocks and TRXs experience different service in network queues\footnote{In Bitcoin, transaction propagation delay is 4 to 5 times larger than blocks \cite{BitcoinCharts}, whereas it is roughly half in Ethereum \cite{Ethna}.}.   
Subsequently, service for block components would also differ from that of TRXs.
For simplicity, we just reflect the priority through service rates, i.e., $\mu _{Ns} =  \zeta \mu _{Nc}$.

We model network as a processor-sharing (PS) queue, since we are dealing with a distributed system with nodes interacting through a shared medium.
When a node in the network receives a piece of new information, it first checks the validity of its content. 
Upon validation, it then relays the information to others, usually by exploiting a variant of gossip protocols \cite{Shahsavari, Ethna}.
Hence, there might be many messages being distributed, relayed and processed simultaneously in the network and different miners can have different perspectives towards what currently is on the network.
Additionally, miners are usually capable enough that can pool their resources among multiple requests.
This leads to a parallel information validation process, also corroborating the PS property.

In particular, we model network as a $./G/1/PS$ queue.
In reality, the information dissemination in the network does not follow an exponential distribution \cite{BitNet, BitcoinCharts}.
However, even if we were to consider exponentially distributed service times for both TRXs and block components in the network (similar to \cite{BlockchainQueue} for the case of TXs), due to the difference between their dissemination strategies ($\zeta \neq 1$ in general) the resultant queue service distribution would not be an exponential.

Regarding the exponential service distribution of PoW mining (see Section \ref{sysModel}), we consider consensus queues as $./M/1/FCFS$. 
Thus, we now have a QN comprised of $./M/1/FCFS$ and $./G/1/PS$ queues connected to each other. 
This QN will then be a PFQN, regarding Poisson arrival of TXs to each shard's network queue. 
The QR property in the queues with non-exponential service times is maintained through symmetric service disciplines \cite{Miyazawa}, e.g., processor sharing network queues in our case.
The proposed QN model then falls within the category of BCMP networks\footnote{Introduced by Baskett, Chandy, Muntz, and Palacios.}, which are one of the predetermined PFQNs \cite{Miyazawa}.

Every time a block is produced, at most $b$ TRXs are removed from the consensus queue and added to the network queue as block components. 
Just like the example in Fig. \ref{fig:Example}, we make use of positive and negative signals to model batch movements.
In doing so, routing probabilities help us to keep track of batch size.
Specifically, by service completion and departure of leading customer in consensus queue (FCFS discipline), it turns into a positive signal with $r_{Pc,N{s _b^ +}} = 1$ and enters the network queue. 
Upon which, the positive signal adds a block component $s$ to the network queue and triggers an entity of its kind at the queue's output.
Every triggered positive signal at the output of the network queue is then routed to consensus queue as a negative signal till $b$ customers are deducted from it and added to network queue. 
In other words, $r_{N{s _i^ +},P{s _{i-1}^ -}} = 1$ and $r_{P{s _{i-1}^ -},N{s _{i-1}^ +}} = 1$ for $i=b, b-1, ..., 2$ and $r_{N{s _1^ +},0} = 1$. 
Hence, every removal of TRX (class $c$ customer) from consensus queue is followed by addition of a block component (class $s$ customer) to the network queue of shard, till either the batch size $b$ is completed or the negative signal meets an empty consensus queue (equivalent to partial batches).

As mentioned before, in order to preserve the QR ‌property, we require network queues to emit positive signals whenever they do not contain block components. 
This additional departure rate is an obvious deviation from the real scenario which is not desirable.
Therefore, as we proposed in case of Fig. \ref{fig:Example}b, we readjust positive signals' rate using routing probabilities.
In particular, we set $r_{N{s _{i+1}^ +},P{s _{i}^ -}} =  {\rho _{N{s}}}$.
This modification does not damage the QR property, though it causes random block sizes.
Nonetheless, it does not interfere with our throughput computation, since rates remain unchanged with respect to the real scenario.

We can now easily write traffic equations for this queueing network.
Due to symmetric architecture, we need to write equations for just one shard, i.e., for just two queues, a consensus queue and its associated network queue in the shard.
Consensus queue traffic equations are  
\begin{subequations} \label{eqConsensus}
	\begin{align}
	\alpha _{\sml{Pc}} & = {\rho _{\sml{Nc}}} {\mu _{\sml{Nc}}},  \label{eqConsensus1} \\
	\alpha_{\sml{Ps_{i}}}^ - & = {\rho^{\sml{-1}} _{\sml{Ns}}} \alpha _{\sml{Ns_{i+1}}}^ + r_{\sml{N{s _{i+1}^ +}},\sml{P{s _{i}^ -}}} = \alpha _{\sml{Ns_{i+1}}}^ +,  \quad \: \: i = 1, ..., b-1. \label{eqConsensus2}
	\end{align}
\end{subequations}
Now, from \eqref{eqConsensus1} and \eqref{eqConsensus2} consensus queue utilization can be easily computed as $\rho _P = {\alpha _{Pc}} (\mu _P + \sum_{i=1}^{b - 1} \alpha _{Ps_{i}}^ -)^{-1}$. 
In the case that TRX fees were also important, we can approximate service policy in consensus queues to be in random order. 
TRX fees are random, hence, miners choices for composing a block would be random.
It does not have any impact on traffic equations, but, one can consider random effect for negative signals in order to capture the effect of TRX fees, i.e., $\eta(l, n) = 1/n$ for $l = 1, ..., n$.

For the network queue, the arrivals are comprised of newly issued transactions (with rate $\lambda$) and relayed receipts in addition to blocks.
That is why its utilization factor, $ {\rho _{N}} $, is comprised of two terms, one for TRXs ($ {\rho _{Nc}} $) and one for block components ($ {\rho _{Ns}} $). 
Network queue traffic equations are
\begin{subequations} \label{eqNet}
	\begin{align}
	\alpha _{\sml{Nc}} & = \lambda  + (M - 1) {\rho _{\sml{Ns}}} {\mu _{\sml{N{s}}}} {r_{\sml{N{s},Nc}}},  \label{eqNet1} \\
	\alpha _{\sml{Ns_{b}}}^ + & = \rho_{\sml{P}} {\mu _{\sml{P}}}, \label{eqNet2} \\
	\alpha _{\sml{Ns_{i}}}^ + & = {\rho _{\sml{P}}} \alpha _{\sml{Ps_{i}}}^ -, \quad \: \: i = 1, ..., b-1. \label{eqNet3}
	\end{align}
\end{subequations} 
Note that each positive input signal ($s _{i}^ +, i = 1, ..., b$), adds a customer of an identical class, $s$, to the network queue.
Obtaining signals' rates in \eqref{eqNet2} and \eqref{eqNet3} is straightforward.
However, obtaining class $c$ customer arrival rate in (\ref{eqNet1}) needs more effort. 
This class represents TXs along with the incoming RXs (due to cross-shard TXs) from the other $M - 1$ shards.
The complication is due to RXs rate. 
In particular, we need to know which block components are meant for other shards, i.e., $r_{N{s},Nc}$ in \eqref{eqNet1}.

Block components represent either new transactions or received receipts from the other $M - 1$ shards. 
Clearly, served RXs must leave the system, but, TXs either become RXs intended for other shards or they are finalized within their shards and exit the system.
Therefore, we first need to know how many components of a block are newly issued TXs, since these are the only ones which may stay in the system.
In \eqref{eqNet1}, the first term is newly issued TXs and we know from \eqref{eqConsensus1} that class $c$ customers in the network queue are directly routed to the consensus queue to be mined.
Hence, each block component $s$ could be stemmed from a newly issued TX with probability $\lambda/\alpha _{Nc}$. 
The number of possible destinations each component can have, due to address-based sharding (uniformly distributed jobs), is $M$.
One for the case sender and receiver addresses belong to the same shard, resulting in local finalization of the TX, and $M - 1$ for the other shards it can be routed to as a RX.
Thus, we get $r_{N{s},Nc} = \frac{1}{M} \frac{\lambda}{{\alpha _{Nc}}}$.

Finally, from \eqref{eqNet} we can easily derive network queue utilization and its components; ${\rho _{N}} =  {\rho _{Nc}} + {\rho _{N{s}}}$ where $\rho _{Nc} = {\alpha _{Nc}} \mu _{Nc}^{-1}$ and $\rho _{N{s}} = (\sum_{i=1}^{b} \alpha _{Ns_{i}}^ +) \mu _{Ns}^{-1}$.

\subsection{Maximum Stable Throughput} \label{shardingThroughput}

Rate stability is the limit for which a queue becomes too crowded so that the jobs stuck in it forever \cite{Kleinrock1}.
In other words, incoming jobs experience infinite delay in an unstable queue. 
This happens when job arrival rate is faster than the service rate, causing jobs to pile up.
In fact, in a lossless queueing system, utilization factor ($\rho$) is defined as the ratio of the incoming rate of jobs to their maximum service rate \cite{Kleinrock1}.
Hence, $\rho < 1$ determines the rate stability region, so that exceeding it, results in system overflow.

We have two queues in our model.
We actually have many more than two.
Nonetheless, due to symmetry we restrict our attention to two typical queues which comprise a shard, a consensus queue and its corresponding network queue.
If a queue of each type becomes unstable, others of the same type will do as well. 
It might seem network queues are more crowded than consensus queues since they host blocks in addition to TRXs. 
However, in reality, network is much faster than consensus.
Take Bitcoin for example, blocks are produced every ten minutes, though it takes couple of seconds to propagate a block \cite{BitcoinCharts}.
Accordingly, we adopt the same approach here by considering $min(\mu_{Nc}, \mu_{Ns}) >> b\mu_P$.
This further limits the occurrence of parallel blocks and thus results in negligible fork rates.

We have all the necessary equations in (\ref{eqConsensus}) and (\ref{eqNet}). 
We can now easily obtain the system throughput.
Solving (\ref{eqNet1}) we get 
\begin{equation} \label{eqTHR}
\lambda = \frac{\alpha_{Nc}}{1 + \frac{M - 1}{M}\frac{\sum_{i=1}^{b} \alpha _{N{s_i}}^+}{\alpha_{Nc}}} = \frac{\alpha_{Nc}}{1 + \frac{M - 1}{M}},
\end{equation}
where the last equality is due to 
\begin{equation} \label{sumPSig}
\sum_{i=1}^{b} \alpha _{N{s_i}}^+ = \rho_P (\mu_P + \sum_{i=1}^{b-1} \alpha _{P{s_i}}^-) = \alpha _{Pc},
\end{equation}
and the fact that $\alpha_{Nc} = \alpha _{Pc}$, since every regular customer of class $c$ in the network queue is ultimately routed  to consensus queue to become finalized. 

Now, substituting $\alpha _{P{s_i}}^-$s in \eqref{sumPSig} with their values from \eqref{eqConsensus2} for $i = 1, 2, ..., b-1$ we will have
\begin{equation} \label{sumPSigExpand}
\alpha_{Nc} = \alpha _{Pc} = \sum_{i=1}^{b} \alpha _{N{s_i}}^+ 
= \rho_P \mu_P + \rho_P^2 \mu_P + ... + \rho_P^b \mu_P = \frac{(1 - \rho_{\sml{P}}^{b})}{1 - \rho_{\sml{P}}} \rho_{\sml{P}} \mu_{\sml{P}},
\end{equation}
which replacing it with $\alpha_{Nc}$ in \eqref{eqTHR} we obtain 
\begin{equation} \label{eqTHRsingle}
\lambda = \frac{\rho_{\sml{P}} (1 - \rho_{\sml{P}}^{b})}{1 - \rho_{\sml{P}}} \frac{\mu_{\sml{P}}}{1 + \frac{M - 1}{M}}.
\end{equation}
As the final step, let us now compute the maximum stable throughput by setting $\rho_{\sml{P}} = 1$.
We would then get
\begin{equation} \label{eqMaxSingleTHR}
\lambda_{max} = \frac{b\mu_{\sml{P}}}{1 + \frac{M - 1}{M}}
\end{equation}
for the case of only single-destination TXs in the system.
Note that \eqref{eqMaxSingleTHR} just gives us an upper bound on the maximum allowable rate of input transactions.   
Also, remember that $\lambda$ represents the TX input rate to each shard.
To obtain the maximum system throughput, we just need to multiply (\ref{eqMaxSingleTHR}) by the number of shards, $M$.

\section{Extension of the Proposed Model to Support Multi-Destination Transactions} \label{extension}

Let us now take one step further and consider a more general case where transactions might have more than one destination.
We consider $d$ destination fields for each TX with probability $D[d]; d \geq 1$, and regard $d_{max}$ as its maximum. 
As the number of destinations in a TX increases, there is a higher chance that it affects more shards.
As we mentioned earlier in Section \ref{sysModel}, other than the destination fields that reside in the same shard as the TX sender, the originating shard produces as many receipts as the number of distinct shards in TX's remaining destination fields.
Hence, we just need to find a way to produce more RXs upon service completion of TXs in their originating shards.

In order to do so, we make use of positive signals in network queues once again.
The difference though is that this time we use them to add receipts to network queues.
We can simply consider addition of a class $c$ customer to the destination shard upon the arrival of the corresponding new positive signals.
In the case of single-destination TXs, we used to simply route a class $c$ customer to the destination shard's network queue.
However, with multi-destination TXs, more RXs are produced and hence, more shards are affected.
With signals, we can use newly triggered signals to visit other shards.
Consequently, we will be able to add the required RXs to the target shards.

Let us clarify the matter with the same example we considered in Section \ref{sysModel}.
A system with three shards $A$, $B$ and $C$, with three TXs issued in shard $A$ with destination shards as `$AA$', `$BB$' and `$BC$'.
Shard $A$ produces none, one and two receipts for each of these TXs, respectively.
In case of a single dispatched RX in `$BB$', we handled it by adding a class $c$ customer to shard $B$ network queue.
For `$BC$' though, we first add a class $c$ customer to shard $B$ by routing a positive signal to its network queue.
Then, we can route the triggered signal at the output of shard $B$ network queue simply as a class $c$ customer to shard $C$ network queue.

Accordingly, we need memory embedded into our signals, since RXs originated from a multi-destination TXs are not allowed to visit a shard twice.
In other words, we need memory to help us keep track of the remaining shards each signal has to still visit.
For our example and the case `$BC$', the triggered signal at the output of shard $B$, needs to know its next target, shard $C$.
We can consider a set-specific positive signal which the set corresponds to the target destination shards.
Upon visit to each of the set members, the newly triggered signal is then routed as (also transformed to) another set-specific positive signal with the leaving shard omitted from the set.
Signals with set of size one are routed to their destination as a regular class $c$ customer.

However, this approach incurs too much complexity.
In fact, we need to introduce routing probabilities for every possible set.
This is not very reasonable due to huge population of sets in case of large $M$ and $d_{max}$.
Instead, we adopt a simpler yet exact strategy.
We drop the notion of a target set in our signals and replace it by stage.
This will allow us to replace the deterministic routing approach with a probabilistic one which is much more straightforward. 
As a result of probabilistic routing scheme, a positive signal might visit a shard more than once, unlike its set-specific counterpart.
Nevertheless, this has no adverse effect on throughput calculation that we are interested in.
Due to uniform load distribution, the incurred load on each queue remains the same, hence the final result would not differ.

Specifically, we introduce multi-stage positive signals where stage represents the number of shards the signal yet to visit.
Let us consider $\mathcal{M} = \{1, 2, ..., M\}$ as the set of all network queues, and $c_{.i}^+$s and $\alpha _{.c_i}^+$s as our multi-stage positive signals and their respective rates, where $i$ represents the stage.
When a multi-stage positive signal enters a network queue, not only does it add a class $c$ customer to the queue, it also triggers a signal of its kind at the output of the queue.
The newly triggered signal is then routed as another multi-stage positive signal with one stage less than its predecessor.
In case the stage of the signal is one, the signal is routed as a regular class $c$ customer.

Let us now go ahead and write the traffic equations for this system.
Since our architecture is still symmetric, the equations are written for one shard, but they apply to others as well.
Also, the changes we make are only restricted to class $c$ customers entering network queues.
Therefore, it only suffices to make some modifications on the corresponding equations of network queues, i.e., \eqref{eqNet1}, and the rest remain intact.
In order to summarize the appearance of equations, we set $c_0^+ = c$ and ${\alpha _{Jc_0}^+} = {\alpha _{Jc}}$.
For a typical network queue $J \in \mathcal{M}$ we would then have 
\begin{equation} \label{eqMultSig}
{\alpha _{\sml{Jc_k}}^+} = \lambda \delta[k] + \sum_{\substack{J' \in \mathcal{M}, \\ J' \neq J}}^{} \left({{\rho _{\sml{J's}}}{\mu _{\sml{J'{s}}}} r_{\sml{J'{s},J{c_k^+}}}} + \rho_{\sml{J'c}}^{-1} \alpha _{\sml{J'c_{k+1}}}^ + r_{\sml{J'{c_{k+1}^+}},\sml{J{c_k^+}}}\right),
\end{equation}
for $k = 0, 1, ..., U$ and $\alpha _{.c_k}^ + = 0$ for all $k > U$, where $U$ is the largest stage possible for a signal and  $\delta[.]$ is a dirac delta function defined over a discrete domain.
A newly generated signal can at most affect either $M - 1$ or $d_{max}$ shards.
Because the last trace of signals are in the form of a class $c$ (equivalently $c_0^+$) customer, then $U = min(M - 1, d_{max}) - 1$.

A stage $k$ signal is either derived from service completion of a receipt or a stage $k+1$ signal. 
From the expression within the parenthesis in \eqref{eqMultSig}, we show the first term with 
\begin{equation} \label{rootEq}
R_{\sml{J'J}} ^k \triangleq {{\rho _{\sml{J's}}}{\mu _{\sml{J'{s}}}} r_{\sml{J'{s},J{c_k^+}}}},
\end{equation}
which is the rate at which service completion of a receipt in network queue $J'$ results in a stage $k$ signal aimed for network queue $J$. 
We will talk more in depth about $R_{J'J} ^k$ and its elements later. 
However, let us first concentrate on the second term within the parenthesis in \eqref{eqMultSig}, referring to higher stage signal rate.

A positive signal can immediately be routed to any shard except the one it is stemmed from.
Hence, due to uniformly distributed routing probabilities, it can be routed to any of the other $M-1$ shards with equal probability.
Accordingly, the second term in the summation of (\ref{eqMultSig}) can be reduced as
\begin{equation} \label{eqRoute}
\rho_{\sml{J'c}}^{-1} \alpha _{\sml{J'c_{k+1}}}^ + r_{\sml{J'{c_{k+1}^+}},\sml{J{c_k^+}}} 
=\rho_{\sml{J'c}}^{-1} \alpha _{\sml{J'c_{k+1}}}^ +  (\frac{\rho_{\sml{J'c}}}{M-1})  = \frac{\alpha _{\sml{J'c_{k+1}}}^ +}{M-1},
\end{equation} 
where $\rho_{J'c}$ is multiplied to prevent the additional departure rate (see Section \ref{analyticModel}).
As mentioned, with probabilistic routing, repetitive visits to a shard by signals are plausible, nevertheless, it does not interfere with our objective.

Substituting \eqref{rootEq} and \eqref{eqRoute} into \eqref{eqMultSig} we get
\begin{equation} \label{eqAlphaKnred}
{\alpha _{\sml{Jc_k}}^+} = \lambda \delta[k] + \sum_{\substack{J' \in \mathcal{M}, \\ J' \neq J}}^{} \left( R_{\sml{J'J}} ^k + \frac{\alpha _{\sml{J'c_{k+1}}}^ +}{M-1} \right),
\end{equation}
for $k = 0, 1, ..., U$ and $\alpha _{.c_k}^ + = 0$ for all $k > U$.
Due to symmetry, each shard equally hosts the same rate of multi-destination TXs as others.
Hence, both rates in summation of \eqref{eqAlphaKnred} are independent of their originating queues.
Therefore, we can simply replace the subscript $J'$ with $J$ in $\alpha _{\sml{J'c_{k+1}}}^ +$ and drop it completely from $R_{\sml{J'J}} ^k$ to get
\begin{equation} \label{eqAlphaK}
{\alpha _{Jc_k}^+} = \lambda \delta[k] + (M-1) R_{J} ^{k} + {\alpha _{Jc_{k+1}}^+},
\end{equation}
for $k = 0, 1, ..., U$ and $\alpha _{.c_k}^ + = 0$ for all $k > U$.
Now, only remains the computation of $R_{J} ^{k}$, the rate a serviced block component $s$ produces a stage $k$ signal.

Blocks are comprised of both RXs and TXs.
The TXs can further be divided into different classes based on the number of their destination fields.
Hence, we need to distinguish the components forming a block to find which can become a positive signal.
In order to do so, we show block components by $s_d, d = 1, ..., d_{max}$ for TXs with $d$ destinations and $s_0$ for RXs.
It is clear that $s_0$ components leave the system after service completion by network queue. 
However, $s_d$ with $d \geq 1$ may or may not leave the system depending on the difference between destination shards and the source shard.
With the newly introduced classes for block components, we can simply rewrite \eqref{rootEq} as 
\begin{equation}\label{eqRoot}
R_{J} ^{k} = {{\rho _{J's}}{\mu _{J'{s}}} \sum_{d=k+1}^{d_{max}} {Pr(s = s_d) r_{J'{s_d},J{c_k^+}}}}.
\end{equation}
The idea is simple.
Upon service completion of a block component, we first check if it is a TX, and if it is, of which class.
Then, the TX is transformed to a positive signal aiming other shards via routing probabilities.
Note that the classification here is just a virtual notion.
We are just using the probability of a block component to have one of the aforementioned classes. 

There are still two unknowns in (\ref{eqRoot}) which we need to find, class probability and its corresponding routing probability.
Let us start with the former.
Following what we had in (\ref{eqNet1}), it is not very complicated to derive the probability corresponding to transactions with $d$ destinations in each block, i.e., for $1 \leq d \leq d_{max}$
\begin{equation} \label{sdClasses}
Pr(s = s_d) =  \frac{\lambda D[d]}{\sum_{k = 0}^{U} {\alpha _{Jc_k}^+}}.
\end{equation}
Basically, the class $c$ customers in a network queue are comprised of both newly issued TXs and RXs routed from other shards.
Hence, the probability of a block component to be a TX can be derived as the ratio of the rate of newly issued TXs to the rate of all class $c$ customers in the network queue.
These TXs can then be categorized with $D[d]$ to have $d$ destination fields.

Nonetheless, the number of destinations in a transaction is not sufficient to obtain the derived signal's stage.
Jobs in our system model are distributed uniformly, therefore, destination fields in a TX might point to the same or different shards.
Recall that we pack pointers to the same shard into a single receipt.
Hence, the positive signal derived from service completion of $s_d$ will usually impact less than $d$ shards.
So, we need probabilities corresponding to what would become of $s_d$ upon service completion.
This is actually handled through routing probabilities $r_{.{s_d},.{c_k^+}}$ in (\ref{eqRoot}). 
While it takes care of the transformation to positive signals, it also has the role of distributing them among shards.

In order to obtain the routing probabilities, the first step is to find the number of distinct shards other than the source shard a multi-destination TX points to.
To clarify, consider once again our example with three shards, $d_{max} = 2$ and shard $A$ producing the TXs.
For TXs with two destination fields, there are $3\times3=9$ possible destination sets.
Among which, one is handled internally (`$AA$').
Six are handled through simple receipts, corresponding to two destinations both pointing to the same shard $B$ or $C$ (`$BB$' and `$CC$'), or cases with one still pointing to shard $A$ while the other points either to shard $B$ or $C$  (e.g., `$AC$').
The remaining two (`$BC$' and `$CB$') correspond to the situation where we need to use positive signals.

To enumerate sets with distinct shards in destination fields of $s_d$, we make use of Stirling number of the second kind \cite{StirlingWebsite}.
The details about how exactly we obtain the mentioned sets population are presented in Appendix \ref{appendixB}.
Here, only the end result suffices for our purpose.
The number of sets with $i$, $i \leq d$, distinct shards other than the originating shard in destination fields of $s_d$ are
\begin{equation} \label{eqCoeff}
N(M,d,i)=\frac{(M - 1)!}{(M - i - 1)!}\stirling{d+1}{i+1},
\end{equation}
where 
\begin{equation} 
\stirling{n}{q} = \frac{1}{q!}\sum_{p=0}^{q}{(-1)^p{q \choose p}(q-p)^n},
\end{equation}
is the Stirling number of the second kind which is the number of ways to partition a set of $n$ objects into $q$ non-empty subsets \cite{StirlingWebsite, concreteMath}.

As mentioned, routing takes care of both transformation and distribution.
The former is obtained by dividing $N(M,d,i)$ in \eqref{eqCoeff} to $M^d$ possible destination sets for $s_d$.
The latter is simply uniform.
So, for \eqref{eqRoot} we get
\begin{equation} \label{eqRk1}
R_{N} ^{k} = \frac{{\rho _{Ns}}{\mu _{Ns}}}{M-1} \sum_{d = k+1}^{d_{max}} Pr(s = s_d) \frac{ N(M,d,k+1) }{M^d},
\end{equation}
where division to $M - 1$ is due to the population of the target shards the newly emerged signal can route to. 
Furthermore, subscript $N$ has been used due to symmetry and identical properties of network queues in $\mathcal{M}$. 
Also, recall that the last trace of a positive signal, $c^+_0$, is a mere class $c$ customer, hence, $k + 1$ distinct shards other than the originating shard results in a $c ^ + _{k}$ signal.
Now, replacing $Pr(s = s_d)$ from \eqref{sdClasses} and deriving $N(M,d,k+1)$ from \eqref{eqCoeff} we get 
\begin{equation} \label{eqRk2}
R_{N} ^{k} = \frac{{\rho _{Ns}}{\mu _{Ns}}}{M-1} \frac{\lambda}{\sum_{k = 0}^{U} {\alpha _{Nc_k}^+}} \sum_{d = k+1}^{d_{max}} \stirling{d+1}{k + 2}  D[d] \frac{ \prod_{z = 1}^{k+1}(M - z)}{M^d},
\end{equation}
for $k = 0, 1, ..., U$, as the rate a serviced block component $s$ produces a stage $k$ signal.

Despite the complicated looks on \eqref{eqRk2}, it is actually pretty easy to derive the throughput.
Starting to solve (\ref{eqAlphaK}) from $k=U$ down to $k=0$, we can obtain the total input rate of class $c$ customers to a network queue as
\begin{equation} \label{eqAlphaK2}
\sum_{k = 0}^{U} {\alpha _{Nc_k}^+} = \lambda  + (M-1) \sum_{k=0}^{U} (k + 1) R_{N} ^{k}
\end{equation}
which then by substituting $R_{N} ^{k}$ with \eqref{eqRk2} we can obtain 
\begin{equation} \label{eqGenTHR0}
\lambda = \frac{\sum_{k = 0}^{U} {\alpha _{Nc_k}^+}}{1 + \frac{{\rho _{Ns}}{\mu _{Ns}}}{\sum_{k = 0}^{U} {\alpha _{Nc_k}^+}} \sum_{k=0}^{U} (k+1) \sum_{d = k+1}^{d_{max}} \stirling{d+1}{k + 2} D[d] \frac{ \prod_{z = 1}^{k+1}(M - z)}{M^d} },
\end{equation}
as the input rate to each shard.
From \eqref{sumPSig} we know that $\sum_{i=1}^{b} \alpha _{N{s_i}}^+  = {\rho _{Ns}}{\mu _{Ns}} = \alpha _{Pc}$.
On the other hand, $\sum_{k = 0}^{U} {\alpha _{Nc_k}^+} = \alpha_{Pc}$ since every regular customer of class $c$ in the network queue is ultimately routed to consensus queue.
Hence, $\sum_{k = 0}^{U} {\alpha _{Nc_k}^+} = {\rho _{Ns}}{\mu _{Ns}}$.
Now, substituting $\sum_{k = 0}^{U} {\alpha _{Nc_k}^+}$ in \eqref{eqGenTHR0} with $\alpha_{Pc}$ from \eqref{sumPSigExpand}, we will have 
\begin{equation} \label{eqGenTHR}
\lambda = \frac{\rho_{\sml{P}} (1 - \rho_{\sml{P}}^{b})}{1 - \rho_{\sml{P}}} \frac{\mu_{\sml{P}}}{1 + \sum_{k=0}^{U} (k+1) \sum_{d = k+1}^{d_{max}} \stirling{d+1}{k + 2} D[d] \frac{ \prod_{z = 1}^{k+1}(M - z)}{M^d}}.
\end{equation}

When $M$ is large enough, the terms with $\stirling{d + 1}{d + 1}$ coefficient become dominant.
They happen when $k+1=d$.
For Stirling number of the second kind $\stirling{n}{n} = 1, \forall n$.
Also, as in Section \ref{shardingThroughput}, we consider $min(\mu_{Nc}, \mu_{Ns}) >> b\mu_P$ so that network is fast enough that does not become the bottleneck.
Accordingly, by putting $\rho_{\sml{P}} = 1$ in \eqref{eqGenTHR} we obtain the maximum throughput per shard as
\begin{equation} \label{eqGenMaxTHR}
\lambda_{max} =  \frac{b \mu_{\sml{P}}}{1 + \sum_{d = 1}^{U + 1} d D[d] \frac{ \prod_{z = 1}^{d}(M - z)}{M^d}}, 
\end{equation}
for sufficiently large $M$.
For a system with a huge number of shards and $d_{max} < M$ we have
\begin{equation} \label{eqFinGenMaxTHR}
\lim_{M \to \infty} \lambda_{max} = \frac{b \mu _{\sml{P}}}{1 + E[d]},
\end{equation}
as the maximum input rate of multi-destination TXs to each shard before the system becomes unstable.

\section{The Computation Sharding Scenario} \label{Mere}

Till now, we have always considered a fully sharded blockchain where not only computation but also storage and relaying tasks are divided among shards.
However, there is also the computation sharding scheme alone which is quite popular among sharding proposals, especially due to its simplicity.
Elastico \cite{Elastico}, uses this approach. 
Even with Monoxide \cite{Monoxide} in its safest mode where all miners operate in all shards, the system can boil down to a computation sharding. 
Nonetheless, we are particularly interested in this sharding scheme since it can serve as a benchmark to illustrate the effect of sharding domains.

\begin{figure}[t!]
	\centering
	\includegraphics[width=0.8\linewidth]{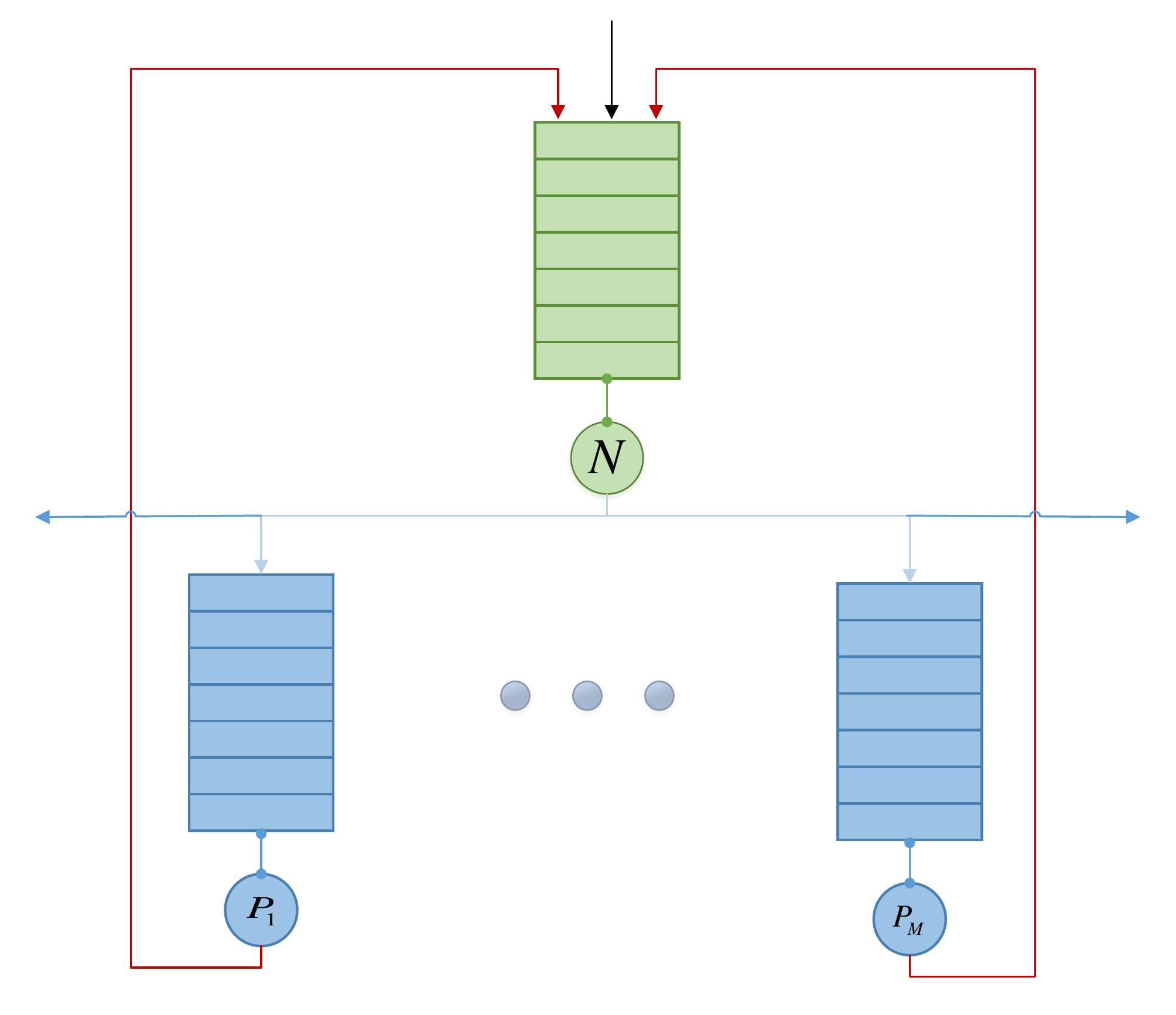}
	\caption{Computation sharding queueing network model. Newly issued transactions are illustrated with black and blocks with red. Blue arrows show committed and finalized TRXs exiting the system.}
	\label{fig:Model3}
\end{figure}

In the computation sharding protocol only the mining task is divided among shards.
All the information is relayed and stored by all the nodes in the network. 
In particular, though every piece of information is broadcast to and stored by the entire network, mining and block production is restricted only to the information related to the corresponding shard.
Fig. \ref{fig:Model3} demonstrates our analytical model modified for this scenario, i.e., a network queue shared among $M$ consensus queues.
As before, each consensus queue produces blocks independent of others and broadcasts it to the whole network for state update.

Since the network is common among all consensus queues, there is no need for a cross-shard communication scheme. 
Receipts are embedded inside blocks that are broadcast and unlike the previous scenario, there is no need for inter-shard re-transmissions.
Thus, there will be no explicit sign of receipts in this model. 
Once a miner receives a block from other shards, it'll be aware of the RXs that are supposed to be added next in its shard chain.
Destination shard may then further postpone the mining process of newly issued RXs (extracted from blocks) for security purposes, i.e., until they are in sufficient depth (dependent on the application) of their shard chain.

Let us now write the traffic equations for this scenario.
Once again, we begin with single destination TXs, i.e., $D[d=1]=1$.
The traffic equations in \eqref{eqConsensus} and \eqref{eqNet} do not actually change much.
In fact, the main differences happen in \eqref{eqConsensus1} and \eqref{eqNet1} when others have only $M$ and $1/M$ multiplied to their right-hand sides for the network queue and consensus queues equations, respectively. 
The main difference with the model in Fig. \ref{fig:Model3} is that the network queue as a shared medium is now responsible for the distribution of loads among consensus queues.
These loads include newly issued TXs and embedded RXs in the blocks.
Accordingly, the aforementioned $M$ and $1/M$ multiplications are respectively due to load aggregation in the network queue and their uniform distribution among consensus queues.

Beginning with consensus queues, we have
\begin{equation} \label{eqMereConsensus1}
{\alpha _{\sml{Pc}}} = {\rho _{\sml{Nc}}} {\mu _{\sml{Nc}}} {r _{\sml{Nc,Pc}}} + {\rho _{\sml{Ns}}} {\mu _{\sml{Ns}}} {r _{\sml{Ns,Pc}}},
\end{equation}
where now ${r _{\sml{Nc,Pc}}} = 1/M$, since network is a shared pool which its content belong to each shard with equal probability.
Also, we have ${r _{\sml{Ns,Pc}}} = \frac{1}{M} \frac{M - 1}{M} \frac{\lambda}{\alpha _{\sml{Pc}}}$ as the share of block components that become RXs to be finalized where $\frac{M - 1}{M}$ is the share of block components not produced by our typical consensus queue. 
 
As for the network queue, we simply have 
\begin{equation} \label{eqMereNet1}
{\alpha _{\sml{Nc}}} = M \lambda, 
\end{equation}
where $\lambda$ is still TX input rate for each shard which becomes of the form $M\lambda$ when aggregated.
One can now easily see from \eqref{eqMereNet1} along with \eqref{eqMereConsensus1}, the inter-shard information propagation through shared network using blocks.
Instead of retransmission of block components to other network queues in a full sharded blockchain, RXs as block components are now directly routed from the network queue to consensus queues of their target shards (second term of \eqref{eqMereConsensus1}).

Solving the traffic equations, once again we obtain the same result as \eqref{eqTHRsingle} for consensus queues throughput rate.
In case of the network queue, one could deduce that service rates would most probably be smaller than that of full sharding case since the network is now comprised of more nodes.
Accordingly, information dissemination will take more time. 
However, we still consider that the network queue service rates satisfy $min(\mu_{Nc}, \mu_{Ns}) >> b \mu_P$.
Nevertheless, this is not enough.
The network queue is more congested here since the load in it increases with the number of shards.
Hence, there is now a chance that network queue could become the bottleneck, though not necessarily the stability bottleneck.
From the security perspective, approaching instability in network queue is intolerable.
Increase in block delivery delay, increases fork rate which can ultimately compromise system safety.
In other words, security limits the throughput of this system long before stability does.
Nonetheless, security concerns are out of the scope of this report and we don't want to deal with delay computations.
Instead, we just set $\rho_{N} < \gamma$ where $\gamma < 1$ is the limit that satisfies our security concerns.

To satisfy $\rho_{N} < \gamma$, we require
\begin{equation}\label{eqNetStabilityPre}
\rho_{N} = \rho_{Nc} + \rho_{Ns} = \frac{M \lambda}{\mu_{Nc}} + \frac{\sum_{i=1}^{b} \alpha _{\sml{Ns_{i}}}^ +}{\mu_{Ns}} < \gamma,
\end{equation}
where $\sum_{i=1}^{b} \alpha _{\sml{Ns_{i}}}^ + = M \frac{\rho_{\sml{P}} (1 - \rho_{\sml{P}}^{b})}{1 - \rho_{\sml{P}}} \mu_P$ ($M$ times bigger than \eqref{sumPSigExpand} due to network as a shared medium).
One can see from \eqref{eqNetStabilityPre} that the load in the network queue increases linearly with the number of shards.
With the help of \eqref{eqTHRsingle}, we have $\sum_{i=1}^{b} \alpha _{\sml{Ns_{i}}}^ + = M \lambda (1 + \frac{M -1}{M})$.
Recall that we set $\mu _{Ns} =  \zeta \mu _{Nc}$ in Section \ref{analyticModel}, hence, substituting $\mu _{Nc}$ and solving \eqref{eqNetStabilityPre} for system throughput we get 
\begin{equation}\label{eqNetStability0}
M \lambda < \frac{\gamma \mu_{Ns}}{\zeta + 1 + \frac{M -1}{M}}.
\end{equation}
It is not very complicated to see that the generalization of traffic equations for this model to the case with multi-destination TXs, also results in \eqref{eqGenTHR} as the throughput rate for consensus queues. 
Hence, the generalization of \eqref{eqNetStability0} would lead to 
\begin{equation}\label{eqNetStability1}
M \lambda < \frac{\gamma \mu_{Ns}}{\zeta + 1 + \sum_{k=0}^{U} (k+1) \sum_{d = 1}^{d_{max}} \stirling{d+1}{k + 2} D[d] \frac{ \prod_{z = 1}^{k+1}(M - z)}{M^d}}.
\end{equation}
As it can be seen from \eqref{eqNetStability0} and \eqref{eqNetStability1}, the system throughput for this sharded blockchain is now bounded, highlighting the effect of sharding domains on scalability.

To further simplify the looks on \eqref{eqNetStability1}, let us obtain the result for fairly large $M$s. 
The system throughput would then be 
\begin{equation}\label{eqNetStability2}
M \lambda < \frac{\gamma \mu_{Ns}}{\zeta + 1 + E[d]}.
\end{equation}
Replacing $\lambda$ with \eqref{eqGenTHR} for large $M$s, we can derive the maximum possible number of shards for this system as 
\begin{equation}\label{eqNetNumShards1}
M < \frac{\gamma \mu_{Ns}}{\bar{b}\rho_P \mu_P (1 + \frac{\zeta}{1 + E[d]})},
\end{equation}
where $\bar{b} = \frac{(1 - \rho_{\sml{P}}^{b})}{1 - \rho_{\sml{P}}}$, obtained from \eqref{sumPSigExpand}, is the average size of the blocks that consensus queues produce.

When $\rho_P < 1$, the number of shards, $M$, can grow till the cumulative load in the network queue makes it the bottleneck, i.e., $\rho_N = \gamma$.
In this case, with less load in consensus queues, we can obtain larger $M$s.
On the other hand, when consensus queues become bottleneck, i.e., $\rho_P = 1$ while $\rho_N < \gamma$, there will be an upper-bound on the possible number of shards.
Introducing $\mu_{NB} = \mu_{Ns}/b$ as the service rate of complete blocks in the network queue, by setting $\rho_P = 1$ in  \eqref{eqNetNumShards1}, we then obtain
\begin{equation}\label{eqMaxNumShards}
M_{max} =  \left \lfloor \frac{\gamma \mu_{NB}}{\frac{\zeta \mu_{P}}{1 + E[d]} + \mu_{P}} \right \rfloor ,
\end{equation}
as the maximum number of shards allowed for the case when consensus queues overflow before the network queue becomes bottleneck.

It is worth mentioning that
the model presented in Fig. \ref{fig:Model3} can also be used to capture the effect of hierarchical sharding designs.
Ethereum 2.0 \cite{Ethereum2.0} uses a beacon chain in its design which is responsible for coordinating nodes among shards and preserving the system security and consistency.
Beacon chain then hosts the blocks' headers of each shard which contain the attesters' (participants in the voting process) signatures. 
The situation is quite similar to that of Fig. \ref{fig:Model3} except that the network queue is now replaced by a beacon chain.
Shards $C_1, C_2, ..., C_M$ then only send their block headers to the beacon chain.
For simplicity we keep the subscript $N$ as well for the beacon chain here and introduce $\mu_{NH}$ as the rate block headers are serviced in the beacon chain.
It is now easy to see that \eqref{eqMaxNumShards} would simply boil down to $M_{max} =  \left \lfloor \mu_{NH}/\mu_{P} \right \rfloor$ with $\gamma = 1$ for the case of stability bottleneck. 
It should be clear that $M_{max}$ for the hierarchical designs is much higher than that of computation sharding.
In the case of computation sharding, the main load in the network queue, as the first term in the denominator of \eqref{eqMaxNumShards} suggests, is usually due to TXs. 
Hence, absence of TX propagation alongside the fact that $\mu_{NH} \geq \mu_{NB}$, allows the beacon chain to host higher number of shards compared to the network queue in the computation sharding case.

\section{Numerical Results} \label{NumResults}

\begin{figure}[t!]
	\centering
	\includegraphics[width=0.85\linewidth]{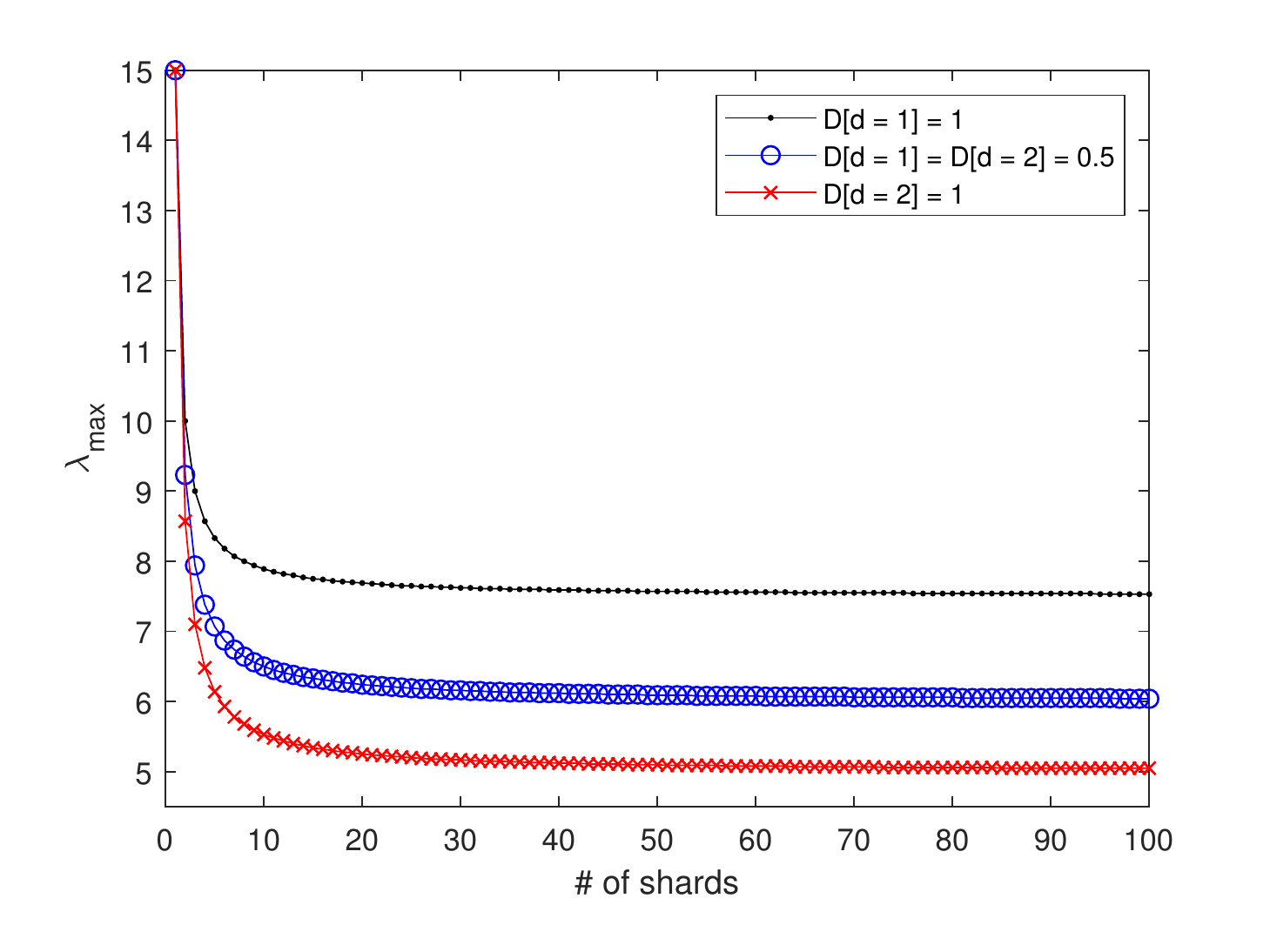}
	\caption{Maximum TX throughput per shard relative to the number of shards in a fully sharded blockchain.}
	\label{fig:maxThroughput}
\end{figure}

In this section, we provide some numerical results to evaluate the maximum stable throughput in different conditions. 
We also show the validity of our analytical model by simulation.
First, we consider the fully sharded blockchain introduced in Section \ref{analyticModel}. 
To derive the maximum stable throughput for this case in Sections \ref{analyticModel} and \ref{extension}, we considered much faster service rate in network queue than consensus queue, i.e.,  $min(\mu_{Nc}, \mu_{Ns}) >> b\mu_P$.
To start with, we set $b = 225$ and $\mu_P = 1/15$, and the maximum throughput per shard $\lambda_{max}$ would change as shown in Fig. \ref{fig:maxThroughput} as the number of shards increases. 
The results are directly stemmed from \eqref{eqTHRsingle} and \eqref{eqGenTHR} for $\rho_P = 1$.
To obtain the system throughput, it suffices to multiply the $x$ and $y$ axis values in Fig. \ref{fig:maxThroughput}. 
Then, it can be seen that there will be a short period of nonlinear growth in system throughput till $\lambda_{max}$ converges and system throughput begins to grow linearly as the number of shards increases.
This is despite the fact that with increase in the number of shards, the load due to wide spread of cross-shard TXs increases.

Also, it can be seen that throughput drops as the number of destinations in TXs increases, which was expected, since TXs now impact more shards. 
However, note that if we were to use single-destination TXs instead of multi-destination TXs, the throughput of the former would have been divided by $E[d]$, highlighting the efficiency of multi-destination TXs.

It is time to support the ideas we exploited in our analytical modeling through simulation.
The objective would then be to find $\lambda$ which saturates consensus queues through simulation and compare it with that of Fig. \ref{fig:maxThroughput}.
We have $N$ shards each comprised of a network and a consensus queue.
TXs arrive at each network queue with rate $\lambda$. 
We uniformly designate each TX a type which defines its destination shard(s).
Consensus queues accumulate TXs and RXs received from other shards to form a block.
If they host more than $b$ TRXs, they produce the biggest block possible. 
Otherwise, the produced blocks would be smaller and equal to the queue length.
After dissemination of the produced block in the network, block RXs leave the system.
However, block TXs are mostly transformed to RXs targeting other shards while some with the same source and destination shards leave the system.
This procedure is performed with the help of TX types.

\begin{figure}[t!]
	\centering
	\includegraphics[width=0.85\linewidth]{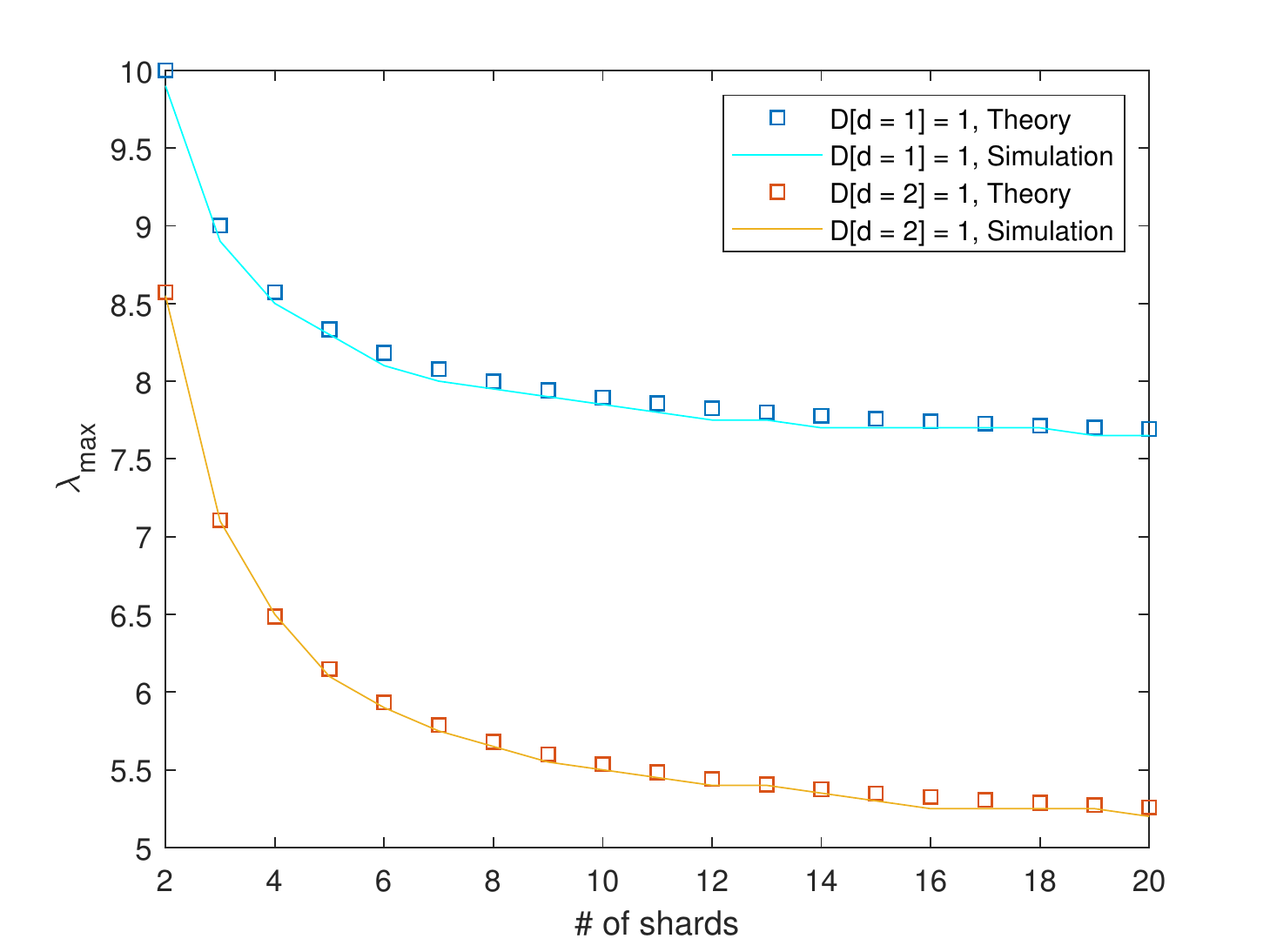}
	\caption{Comparison between theory and simulation results of maximum stable throughput of shards in a fully sharded blockchain.}
	\label{fig:simThroughput}
\end{figure}

Different configurations are then obtained by sweeping $\lambda$ and $M$.
We increase $\lambda$ by $0.05$ steps each time for a specific $M$ till consensus queues start to saturate.
To detect the saturation, it only suffices to check whether total arrivals to a queue exceeds total departures from it in a sufficiently large time interval.
Fig. \ref{fig:simThroughput} shows the comparison of the results from \eqref{eqMaxSingleTHR} and that of simulation for single and double destination TXs.
As can be seen the obtained results are quite close, confirming the validity of our model.

\begin{figure}[t!]
	\centering
	\includegraphics[width=0.85\linewidth]{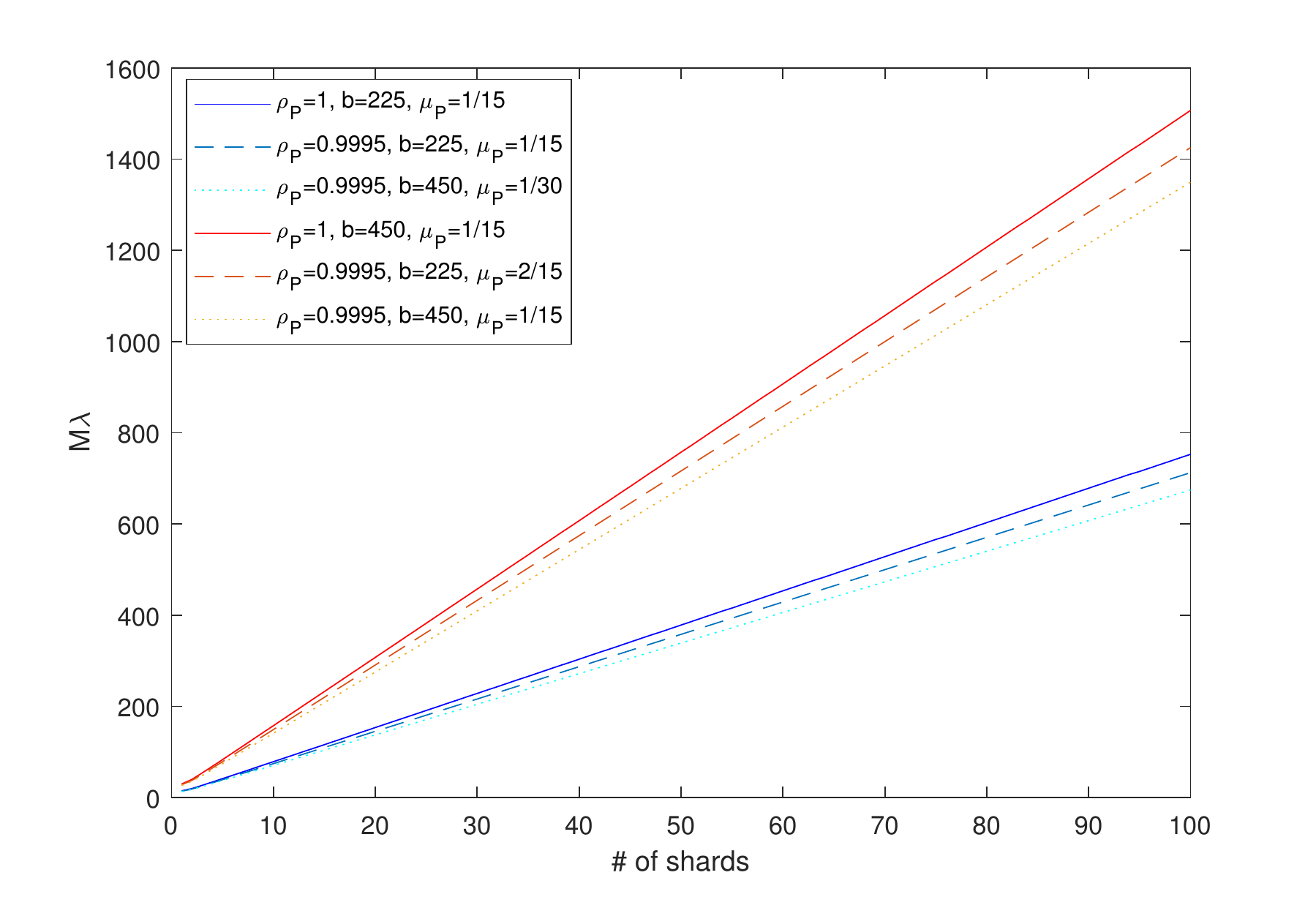}
	\caption{System throughput for $\rho _P = 0.9995$ and different values of $b$ and $\mu_P$.}
	\label{fig:throughputDynamic}
\end{figure}

Let us also inspect the effect of $b$ and $\mu _P$ separately. 
In order to do so, we set $\rho _P = 0.9995$ in \eqref{eqTHRsingle} to be able to compare the effect of $b$ and $\mu _P$ on $\lambda$ with each other.
The result is shown in Fig. \ref{fig:throughputDynamic} except that instead of $\lambda$, the system throughput, $M \lambda$, is displayed.
It can be seen that doubling the service rate requires higher $\lambda$s than doubling the block size to achieve the same value of $\rho _P = 0.9995$. 
In other words, in order to reduce delay, increasing service rate is more efficient than increasing block size. 
Though, depending on fork rate targets in the system, this might not always be desirable.

Note that though throughput grows linearly with $M$ in Fig. \ref{fig:throughputDynamic}, this isn't necessarily the case in reality.
We are only considering stability as the limit for throughput growth.
There are many other important factors in determining the system throughput, and stability though necessary is not sufficient.
Probably, the most important factor is the system security.
Usually when we add new shards, the number of miners in each shard reduces, leaving shards more susceptible to take-over.
Also, we have disregarded the importance of block delivery delay on the safety of the system.
There are many aspects towards the security of blockchain sharding \cite{Security} which need to be considered when evaluating its performance. 
So, what Fig. \ref{fig:throughputDynamic} shows is an upper-bound on throughput which a fully sharded blockchain can achieve given only that all the other issues are addressed. 
In fact, just recall the scalability trilemma which states that we cannot fully achieve scalability, security and decentralization in a blockchain all together at the same time.

\begin{figure}[t!]
	\centering
	\includegraphics[width=0.85\linewidth]{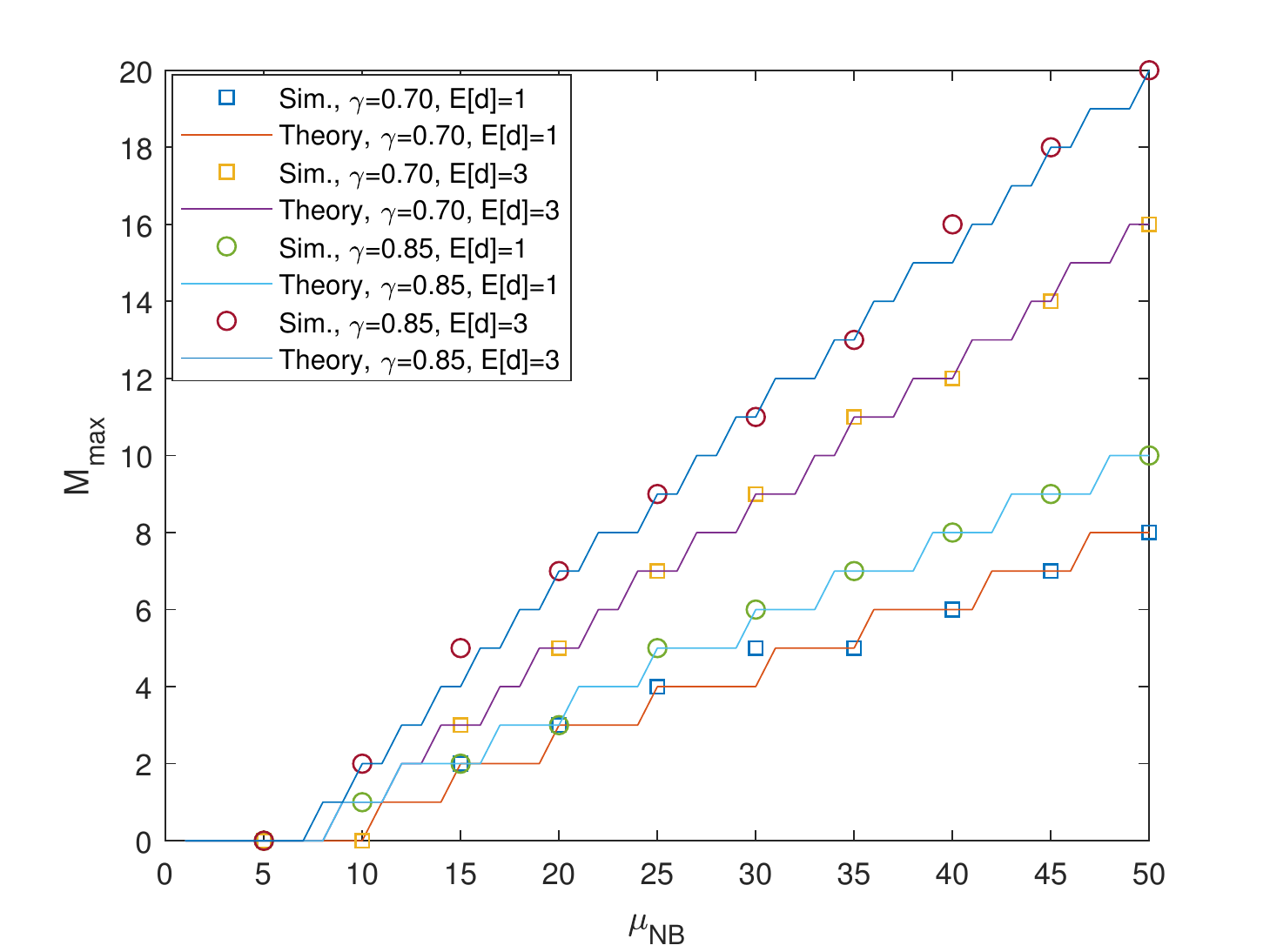}
	\caption{Maximum number of shards possible derived both from theory and simulation for the computation sharding scenario when consensus queues saturate before the network queue utilization reaches $\gamma$.}
	\label{fig:MereThroughput}
\end{figure}

Now, let us evaluate the performance of the computation sharding represented by the model in Fig. \ref{fig:Model3}.
We consider a computation sharding scenario with the set of parametes as $b=225$, $\mu_P=1/15$ and $\mu_{NB} = 0.5\mu_{Nc}$  (equivalently $\zeta = b/2$). 
Fig. \ref{fig:MereThroughput} illustrates the maximum possible number of shards derived both from \eqref{eqNetStability1} and simulation.  
They are obtained for the cases when consensus queues saturate before the network queue reaches $\gamma$.
Since \eqref{eqNetStability1} results in a polynomial equation, one can use \eqref{eqMaxNumShards} as the starting point to find $M_{max}$.
Also, to obtain the utilization factor in simulations, it just suffices to measure the ratio of times that the network queue is busy.

As can be seen from Fig. \ref{fig:MereThroughput}, with single-destination TXs, not that many shards can be achieved. 
The situation improves with multi-destination TXs, since communication overhead reduces due to implicit RXs.
Nonetheless, there is a good chance that $E[d]$ will be small in practice.
Higher $\gamma$s can result in larger $M$s, however, it comes at the cost of higher fork rates.
It is worth mentioning that for this case, since $\rho_P =1$ we have $\bar{b} = b$ and thus $\mu_{NB}$ represents the actual service rate of blocks in the network.

\begin{figure}[t!]
	\centering
	\includegraphics[width=0.85\linewidth]{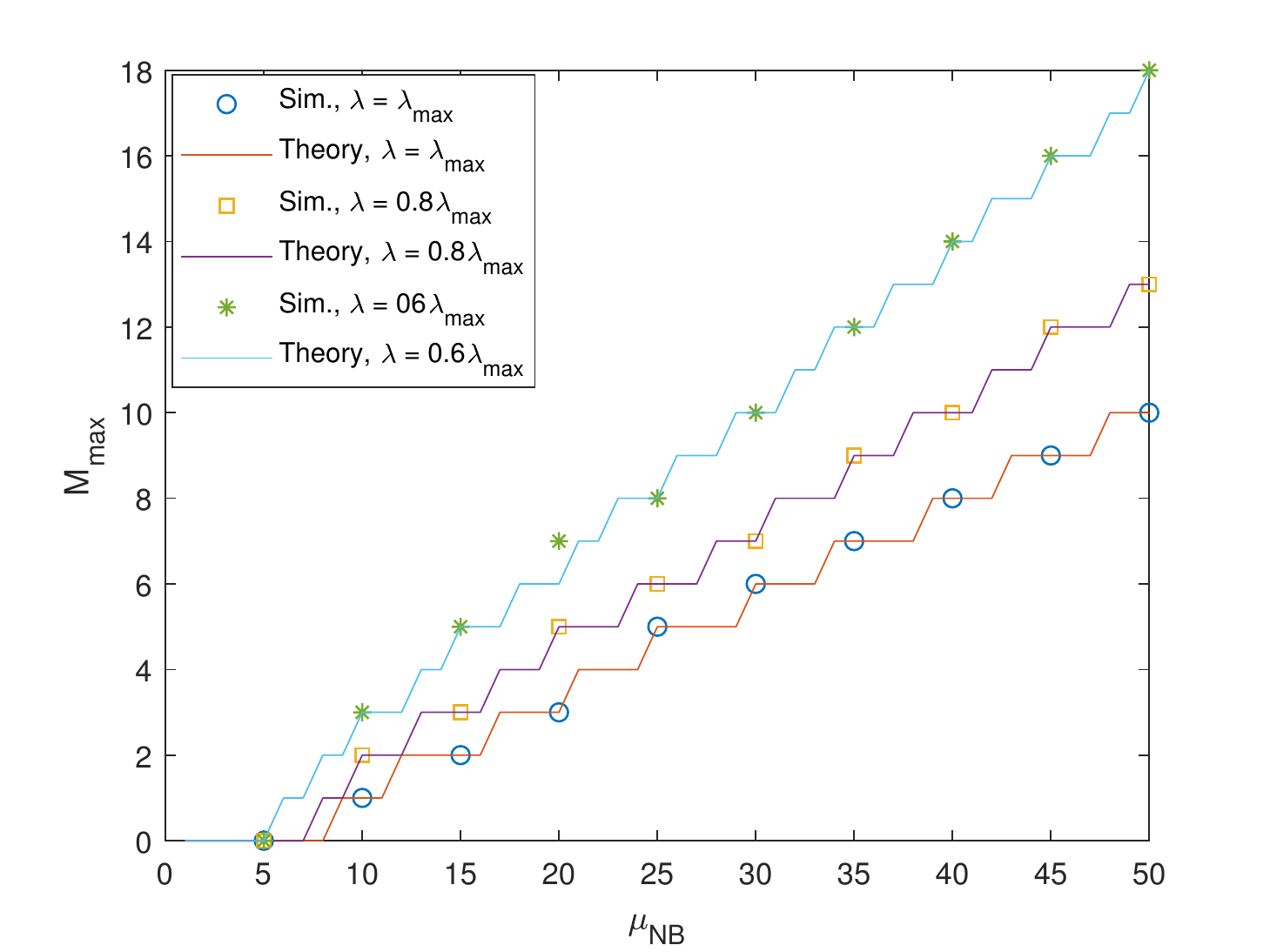}
	\caption{Maximum number of shards possible derived both from theory and simulation for the computation sharding scenario when consensus queues are not necessarily the bottleneck. In this case, $\gamma=0.85$ and $E[d] = 1$.}
	\label{fig:mereThroughputsim}
\end{figure}

When consensus queues are fed less than their saturation limit, Fig. \ref{fig:mereThroughputsim} shows the number of shards we can achieve for the case of $\gamma=0.85$ and $E[d] = 1$.
Clearly, smaller $\lambda$s can result in higher $M$s.
Note that though we decrease $\lambda$ to achieve higher $M_{max}$, the system throughput, $M_{max}\lambda$ for different $\lambda$s remain on the same level as \eqref{eqNetStability1} suggests.
Furthermore, with faster network, we can obtain higher $M$s.
However, there is not usually much control over $\mu_{NB}$ due to decentralized and heterogeneous nature of the network.

In the end, it is worth noting that comparing full sharding with computation sharding, though the total throughput for the latter is limited, it is on the other hand more secure in terms of facing adversary due to more audibility.
This further highlights the importance of scalability trilemma in blockchains.

\section{Conclusion} \label{Conclusion}

In this report, we proposed queueing network models to study the sharding performance in blockchains.
There can be many configurations possible for modeling a sharded blockchain due to the extent of sharding in different domains of computation, network and storage.
In particular, we proposed two models, one for a fully sharded blockchain and one for when only the computation sharding is employed.
In the former miners were exclusive to each shard in terms of their responsibilities, i.e., block production, relaying and storage.
In the latter though, only block production was exclusive and miners relayed and stored every piece of information.
We then obtained closed-from solutions for the maximum stable throughput of these setups.
We explained that the stability is a necessary condition which can further provide us with an upper-bound on throughput in real scenarios. 
We showed that given one can address security issues in a full sharded setup, it can scale with the number of shards.
On the other hand, in the computation sharding setup, we derived the throughput with respect to fork rate limit which the system can tolerate, since stability is no longer relevant.
We showed that in computation shardings the overall throughput cannot grow indefinitely. 
These evaluations brings up the idea that for blockchains to scale more freely, computation sharding should be shipped with network and storage shardings at least to some extent.

\appendices
\section{} \label{appendixA}

In this section, we elaborate quasi-reversibility a bit more in depth. 
It should be noted that the composition of this section is heavily relied on \cite{Miyazawa}.
We begin with introducing the embedded counting processes and specify when they would be a Poisson process.
These are the pillar of QR definition which is closely related to Poisson flows.
Then, we introduce the conditions for queues with signal to be QR.
Finally, we connect QR queues into a network to show that such a network has a product-form solution.
As a result, we can then isolate each queue from the network and examine it individually after solving the traffic equations.

Let us consider a continuous time Markov chain (CTMC) ${X (t)}$ with $q(x,x')$ representing the transition rate from state $x$ to $x'$, for $x,x' \in \mathcal{S}$ where $\mathcal{S}$ is the state space.
Let $N(t)$ now be a counting process that counts the number of state transitions in process $X(t)$ from $x$ to $x'$ with probability $q_*(x, x')/q(x, x')$ independently of other events.
We would then call $q_*$, a thinned transition rate of $q$, i.e., $q_* (x,x') \leq q(x,x')$, and the process $N(t)$ is called an embedded counting process generated by $q_*$ with respect to $q$.
If there exists a positive constant $\lambda$ such that
\begin{equation} \label{embedded}
\sum_{x' \in \mathcal{S}} q_*(x, x') = \lambda, \:\:\:\: for \: all \: x \in \mathcal{S}, 
\end{equation}
then the counting process generated by $q_*$ with respect to $q$ is a Poisson process with arrival rate $\lambda$. 
Accordingly, the future epochs of the embedded counting process would be independent of the current state and the past history of the Markov chain.

Quasi-reversibility is a property concerning the arrival and departure processes.
Therefore, for each pair of states $x,x' \in \mathcal{S}$, we need to decompose the transition rate function $q(x, x')$ of the queue into three types of rates, namely
\begin{subequations} \nonumber
	\begin{align}
	&q_u^A(x, x'), \:\:\:\: u \in T\\
	&q_v^D(x, x'), \:\:\:\: v \in T\\
	&q^I(x, x'),
	\end{align}
\end{subequations}
where $T$ is the union of both arrival and departure classes, which is countable.
These thinned transition rate functions $q_u^A$, $q_v^D$ and $q^I$ generate the embedded point processes corresponding to class $u$ arrivals, class $v$ departures and the internal transitions, respectively. 
Perhaps, only $q^I$ needs a bit more explanation, though it's not particularly of our interest.
An internal transition typically represents a change in status of the customers such as a decrease in their remaining service times.

Before going through the QR definition, we just need to introduce the triggering probability function.
Assume that when a class $u$ entity arrives and induces the state of the queue to change from $x$ to $x'$, it instantaneously triggers a class $v$ departure with probability $f_{u, v}(x, x')$, where
\begin{equation} \nonumber
\sum_{v \in T} f_{u, v}(x, x') \leq 1, \:\:\:\:for\: u \in T, x, x' \in \mathcal{S}.  
\end{equation}
This function particularly enables us to capture the effect of signals.
In this report, only signals can cause triggering and they only trigger their own kind at the departure. 

\begin{definition}
	(quasi-reversibility of queues with signals)
	Assume that $q$ admits the stationary distribution $\pi$.
	Then, if there exist two sets of nonnegative numbers $\{\alpha_u; u \in T\}$ and $\{\beta_u; u \in T\}$ such that
	\begin{subequations} \label{QR-def}
		\begin{equation}
		\sum_{x' \in \mathcal{S}} q_u^A(x, x') = \alpha_u, \:\:\:\: x \in \mathcal{S}, u \in T, \label{QR1} \\
		\end{equation}
		\begin{equation}
		\sum_{x' \in \mathcal{S}} \pi(x') \left(q_v^D(x', x) + \sum_{v \in T} q_v^A(x', x)f_{v, u}(x', x) \right) 
		= \beta_u \pi(x), \:\:\:\:\:\: x \in \mathcal{S}, u \in T \label{QR2} ,
		\end{equation}
	\end{subequations}
	the queue with signals is said to be quasi-reversible with respect to $\{q_u^A, f_{u, v}; u \in T, v \in T\}$, $\{q_u^D; u \in T\}$, and $q^I$.
\end{definition}

One can easily recognize that \eqref{QR1} and \eqref{embedded} are basically the same.
Hence, the first condition of QR property is Poisson arrival of each input entitiy to the queue.  
The second condition is the same as the first only in reverse direction.
It makes sure that the departures also follow Poisson process.
In other words, QR implies that the arrivals of different classes of entities form independent Poisson processes, and the departures of different classes of entities, including both triggered and nontriggered departures, also form independent Poisson processes. 

Perhaps we could elaborate this topic better via an example. 
Consider an $M/M/1$ queue with two classes of arrivals denoted by $c$ and $s^-$, a regular customer and a negative signal, respectively.
When an $s^-$ arrives at the queue, it triggers a customer to depart immediately as a class $s^-$ departure, provided the queue is not
empty upon its arrival. 
If a signal arrives at an empty queue, nothing occurs and no departure is triggered. 
The customer departures generated by regular service completions are still classified as class $c$ departures.
The decomposed transition rates are
\begin{subequations} \nonumber
	\begin{align}
	&q_c^A(n, n+1) = \alpha, \:\:\:\: n \geq 0,\\
	&q_{s^-}^A(n, n-1) = \alpha^-, \:\:\:\: n \geq 1,\\
	&q_{s^-}^A(0, 0) = \alpha^-, \\
	&q_{c}^D(n, n-1) = \mu , \:\:\:\: n \geq 1.
	\end{align}
\end{subequations}
All other transition rates are zero.
Now, according to \eqref{QR1}, the arrivals of regular customers and negative signals are independent Poisson processes with rates $\alpha$ and $\alpha^-$, respectively.
Note that defining $q_{s^-}^A$ for state zero, while it has no effect, is absolutely necessary in order to satisfy \eqref{QR1} and push negative signal under the category of Poisson process.

Let us now check the requirement in condition \eqref{QR2}.
As for the triggering mechanism, we have
\begin{subequations} \nonumber
	\begin{align}
	&f_{c, c}(n, n') = f_{c, s^-}(n, n') = 0, \:\:\:\: n, n' \geq 0,\\
	&f_{s^-, s^-}(n, n-1) = 1, \:\:\:\: n \geq 1.
	\end{align}
\end{subequations} 
We then only need the stationary distributions to derive departure rates from \eqref{QR2}.
Since the dynamics of this queue are the same as those of a regular $M/M/1$ queue with arrival rate $\alpha$ and service rate $\mu + \alpha^-$, its stationary distribution $\pi$ is given by
\begin{equation}
\pi(n) = \left(1 - \frac{\alpha}{\mu + \alpha^-}\right){\left(\frac{\alpha}{\mu + \alpha^-}\right)}^n. 
\end{equation}
Accordingly, for utilization factor we would have $\rho = \frac{\alpha}{\mu + \alpha^-}$.
Now, if we set $\beta = \rho \alpha$ and $\beta^- = \rho \alpha^-$, respectively as for customer and negative signal departure rates in \eqref{QR2}, then this system would be quasi-reversible with departure rates $\beta$ and $\beta^-$.
Accordingly, the departures would also be independent Poisson processes.

We can also include positive signal to this example. 
The positive signal $s^+$ adds a customer to the queue upon its arrival and thus we would have
\begin{equation} \nonumber
q_{s^+}^A(n, n+1) = \alpha^+, \:\:\:\: n \geq 0.
\end{equation}
As for triggering we also consider
\begin{equation} \nonumber
f_{s^+, s^+}(n, n+1) = 1, \:\:\:\: n \geq 0.
\end{equation}
The utilization factor and stationary distribution then would be $\rho = \frac{\alpha + \alpha^+}{\mu + \alpha^-}$ and $\pi(n)=(1-\rho)\rho^n$, respectively.
Now, checking the conditions in \eqref{QR-def}, we would find $\beta^+$, the positive signal departure rate, to be state dependent since $\beta^+=0$ at state zero, while it is $\beta^+ = \rho^{-1} \alpha^+$ for others.
Hence, the queue does not satisfy the QR conditions anymore.
To remedy that, we need to modify the queue to emit a Poisson departure process of positive signals with rate $\rho^{-1} \alpha^+$, i.e., $q_{s^+}^D(0, 0) = \rho^{-1} \alpha^+$.
Consequently, in order to preserve the QR ‌property in queues with positive signals, an additional departure rate is necessary.

Let us now inter-connect $M$ queues that are QR in isolation in an arbitrary manner, comprising a queueing network.
The interactions between the queues are defined as follows.
A class $u$ departure from queue $J$ enters queue $K$ as a class $v$ arrival with probability $r_{Ju,Kv}$. 
We require
\begin{equation}
\sum_{K=0}^{M} \sum_{v \in T_K} r_{\sml{Ju,Kv}} = 1, \:\:\:\:\:\: J = 0, 1, ..., M, \: u \in T_J,	
\end{equation}
where 0 represents the exogenous world.
Class $u \in T_0$ entities arrive at the network from the exogenous world at rate $\beta_{0u}$, which is known, then each entity joins node $K$ as a class $v$ entity with probability $r_{0u,Kv}$.
Let $\beta_{Kv}$ be the average departure rate of class $v$ entities from node $K$. 
The average arrival rate of class $u$ entities at node $J$ satisfies
\begin{equation} \label{trafficEqs}
\alpha_{\sml{Ju}} = \sum_{K=0}^{M} \sum_{v \in T_K} \beta_{\sml{Kv}} r_{\sml{Ju,Kv}}, \:\:\:\:\:\: J = 0, 1, ..., M, \: u \in T_J.	
\end{equation}
These equations are referred to as the traffic equations which are in general nonlinear in $\alpha_{\sml{Ju}}$s.
We always have the notion of traffic equation for queueing networks, but it is only through QR that $\alpha_{\sml{Ju}}$ and $\beta_{\sml{Kv}}$ in \eqref{trafficEqs} are Poisson processes and independent of each other.
This gives us the main result of this section.

\begin{theorem}
	If each queue $J$, $J = 1, ..., M$, with signals is quasi-reversible with $\overrightarrow{\alpha_J}$, $\overrightarrow{\alpha_J} = (\alpha_{Ju}; u \in T_J)$, that is the solution to the traffic equations \eqref{trafficEqs}, then the queueing network with signals has the product-form stationary distribution
	\begin{equation} \label{PFQN}
	\pi(\overrightarrow{\rm x}) = \prod_{J=1}^{M} \pi_J^{(\overrightarrow{\alpha_J})} (x_J) \:\:\:\:\:\:   \overrightarrow{\rm x} = (x_1, ..., x_M) \in \mathcal{S},	
	\end{equation}
	where $\pi_J^{(\overrightarrow{\alpha_J})}$ is the stationary distribution of $q_J^{(\overrightarrow{\alpha_J})}$, $J = 1, ..., M$.
\end{theorem}

\begin{proof}
	See \cite{Miyazawa} chapter 4.
\end{proof}

Hence, in networks comprised of only QR queues, the joint distribution of all the queues is the product of the marginal distributions of the individual queues, provided that the corresponding traffic equations are satisfied.  
Due to this property, the analysis of a queueing network reduces to the analysis of each single queue solely, as well as solving the traffic equations.
In other words, we can now isolate each queue from the network and examine it individually knowing that its arrivals and departures would be Poisson which their rates can easily be derived from \eqref{trafficEqs}.

\section{} \label{appendixB}

In this section we enumerate sets with distinct shards in destination fields of $s_d$.
However, there is a caveat. 
We need to ignore the shard that originates $s_d$ in the process of enumeration since their relative operations are taken care of internally within the originating shard; thus, have no effect on the stage of the signal.
Accordingly, we consider $i$, $i \leq d$, the number of distinct shards other than the originating shard in destination fields of $s_d$.
Then, we just need to find the number of sequences of length $d$, containing exactly $i$ elements from $M$.

First, we assume that there is no trace of the originating shard in destination fields of $s_d$.  
The number of sequences of length $d$ using some $i$ elements, such that each element is used at least once is exactly the number of surjections\footnote{The function $f \colon X \to Y$ is surjective if every element of $Y$ is the image of at least one element of $X$, i.e., $\forall y \in Y, \exists x \in X$ such that $f(x)=y$  \cite{surjection}.} from the set $\{1,2, ...,d\}$ onto the set $\{1,2, ...,i\}$.  
The number of such sequences by the inclusion-exclusion principle is equal to
\begin{equation} \label{Stirling}
N_1(d,i)=\sum_{j=0}^{i}{(-1)^j{i \choose j}(i-j)^d}=i! \stirling{d}{i},
\end{equation}
where $\stirling{d}{i}$ denotes the Stirling number of the second kind which is the number of ways to partition a set of $d$ objects into $i$ non-empty subsets \cite{concreteMath, StirlingWebsite}.
The relation for computing the Stirling numbers of the second kind can be derived from \eqref{Stirling}.
Since we calculated $N_1$ for a fixed set of $i$ elements, we need to consider that these elements can be selected in ${M - 1 \choose i}$ ways (we are excluding the originating shard). So, we have
\begin{equation}
N_2(M, d, i)= {M - 1 \choose i} N_1(d,i) = {M - 1 \choose i} i! \stirling{d}{i}.
\end{equation}

Including the originating shard is not that difficult. 
We know that the originating shard can appear at least $0$ and at most $d - i$ times, since our sequence must have at least one of each unique element.
Also, we can select $l$ places for the originating shard in ${d \choose l}$ ways. 
So, the number of elements in the group $i$ would exactly be
\begin{equation}
N_3(M, d, i)= \sum_{l=0}^{d-i}{d \choose l} N_2(M, d - l, i) = \frac{(M - 1)!}{(M - i - 1)!} \sum_{l=0}^{d-i}{d \choose l} \stirling{d - l}{i}.
\end{equation}
Now, using the property of Stirling numbers as 
\begin{equation}
\stirling{n + 1}{m + 1} =\sum_{j=m}^{n}{{n \choose j} \stirling{j}{m}},
\end{equation}
we obtain
\begin{equation}
N_3(M, d, i) = \frac{(M - 1)!}{(M - i - 1)!} \stirling{d + 1}{i + 1},
\end{equation}
as the number of sets with $i$ distinct shards other than the originating shard in destination fields of $s_d$.
Interested readers can refer to \cite{mathStack} for more information.

\ifCLASSOPTIONcompsoc
  \section*{Acknowledgments}
\else
  \section*{Acknowledgment}
\fi

This publication was supported by grant No. RD-51-9911-0030 from the R\&D Center of Mobile Telecommunication Company of Iran (MCI) for advancing information and communications technologies.

\ifCLASSOPTIONcaptionsoff
  \newpage
\fi

\bibliographystyle{IEEEtran}
\bibliography{IEEEabrv, references}

\end{document}